\documentclass{raa}            % referee version: for submission

\usepackage{graphicx,times,natbib}             %for PS/EPS graphics inclusion, new

\begin{document}

   \title{What determines the observational differences of blazars?
%\footnotetext{$*$ Supported by the National Natural Science Foundation of China.}
}
%   \subtitle{I. Place Your Subtitle Here}

   \volnopage{Vol.0 (200x) No.0, 000--000}      %%preserved for Editor. DOn't remove!
   \setcounter{page}{1}          %%starting page, preserved for Editor. DOn't remove!

   \author{Xu-Liang Fan
      \inst{1,2,3}
   \and Jin-Ming Bai
      \inst{1,2}
   \and Jirong Mao
      \inst{1,2}
   }

   \institute{Yunnan Observatories, Chinese Academy of Sciences, Kunming 650011, China; {\it fxl1987@ynao.ac.cn, baijinming@ynao.ac.cn}\\
        \and
             Key Laboratory for the Structure and Evolution of Celestial Objects, Chinese Academy of Sciences, Kunming 650011, China\\
        \and
             University of Chinese Academy of Sciences, Beijing 100049, China\\
   }

   \date{Received~~2009 month day; accepted~~2009~~month day}

\abstract{We examine the scenario that the Doppler factor determines the observational differences of blazars in this paper. Significantly negative correlations are found between the observational synchrotron peak frequency and the Doppler factor. After correcting the Doppler boosting, the intrinsic peak frequency further has a tightly linear relation with the Doppler factor. It is more interesting that this relation is consistent with the scenario that the black hole mass governs both the bulk Lorentz factor and the synchrotron peak frequency. In addition, the distinction of the kinetic jet powers between BL Lacs and FSRQs disappears after the boosting factor $\delta^2$ is considered. The negative correlation between the peak frequency and the observational isotropic luminosity, known as the blazar sequence, also disappears after the Doppler boosting is corrected. We also find that the correlation between the Compton dominance and the Doppler factor exists for all types of blazars. Therefore, this correlation is unsuitable to examine the external Compton emission dominance.
\keywords{galaxies: jets --- BL Lacertae objects: general --- quasars: general --- radiation mechanisms: non-thermal}
}

   \authorrunning{X.-L. Fan, J.-M. Bai \& J.-R. Mao }            %author_head in even pages
   \titlerunning{Blazar Classification}  % title_head in odd pages

   \maketitle
%% The author head (on even pages) and the title head (on odd pages) will be
%% automatically extracted from \author{} and \title{}. Whenever the title is too long,
%% you will be asked to supply a shorter one by inserting either \authorrunning{} or
%% \titlerunning{} before \maketitle. Anyway, you can specify your own heads.
%%
%%
%% Note: In the following text body of your manuscript, please note several differences from
%%       other major journals:
%% (1) \subsection{Please Capitalize the First Letter of Each Notional Word in Subsection Title}
%% (2) Please Capitalize the First Letter of Each Notional Word in all tables' captions

%
%________________________________________________ sections below
%
\section{Introduction}           %% first-level sections will be auto-capitalized
\label{sect:intro}

Blazar is the most extreme subclass of active galactic nuclei (AGNs). Its radiation is dominated by the non-thermal emission of relativistic jet with small viewing angle to our line of sight. The spectral energy distributions (SEDs) of blazars show two peaks (in $\nu - \nu L_\nu$ diagram) which are believed to be produced by synchrotron and inverse Compton (IC) processes, respectively. However, whether the external photons outside jet participate in the IC process is still an open question (e.g., \citealt{2011ApJ...735..108C}; \citealt{2012ApJ...752L...4M}).

Blazars are classified as flat spectrum radio quasars (FSRQs) and BL Lac objects (BL Lacs) by the optical spectra. They can also be classified as low synchrotron peaked blazars (LSPs), intermediate synchrotron peaked blazars (ISPs), and high synchrotron peaked blazars (HSPs) based on the synchrotron peak frequency~\citep{2010ApJ...716...30A}. An alternative classification based on the ratio of broad line luminosity to Eddington luminosity was proposed by~\citet{2011MNRAS.414.2674G}. This classification considers the potential selection effect on the equivalent width (EW) measurement of broad lines due to the Doppler boosting effect, and links the observational classification to accretion regimes. On the other side, the blazar sequence based on the bolometric luminosity was put forward to unify the observational differences of blazars~\citep{1998MNRAS.299..433F}. The negative correlations between the peak frequency and  luminosity, as well as the Compton dominance (CD) are explained as the increasing cooling from external photons outside jet with increasing luminosity \citep{1998MNRAS.301..451G}. However, latter studies showed that the blazar sequence was an artefact of the Doppler boosting~\citep{2008A&A...488..867N} or redshift selection effects~\citep{2012MNRAS.420.2899G}, and the sources with both high luminosity and high peak frequency which break the blazar sequence exist (e.g., \citealt{2012MNRAS.422L..48P}; \citealt{2015ApJ...810...14A}). To improve the simple blazar sequence, \citet{2011ApJ...740...98M} proposed a concept named ``blazar envelope" which considered the jet power and the orientation effect. According to the blazar envelope, blazars are composed by two populations divided by jet power, and an envelope forms due to different orientation for various sources.

 Since its launch in 2008, the large area telescope (LAT) onboard Fermi gamma-ray space telescope (Fermi) had detected 1444 AGNs in the third LAT catalog (3LAC) clean sample \citep{2015ApJ...810...14A}. The broad energy range and high accuracy of LAT promote deep understandings on both the radiation mechanism (e.g. \citealt{2011ApJ...735..108C}; \citealt{2012ApJ...752L...4M}) and the classification of blazars (e.g. \citealt{2009MNRAS.396L.105G}; \citealt{2011MNRAS.414.2674G}). However, it is still uncertain that the differences of blazars are determined by different physical features, or observational effects (such as the orientation). Moreover, it also needs to be verified that there are one or more factors taking effects on the blazar classifications.

The distinctions of the Doppler boosting had been discovered for different subclasses of blazars, such as BL Lacs and FSRQs \citep{2009A&A...494..527H}, or X-ray-selected BL Lacs and radio-selected BL Lacs \citep{1993ApJ...407...65G}. Stronger Doppler boosting was also suggested to explain the $\gamma$-ray detected blazars by many papers (e.g. \citealt{2009ApJ...696L..17K}; \citealt{2009ApJ...696L..22L}; \citealt{2010A&A...512A..24S}; \citealt{2015ApJ...810L...9L}). Furthermore, the blazar sequence was found to be an artefact of the Doppler boosting \citep{2008A&A...488..867N}. All these results indicate that the observational differences of blazars could be determined by the Doppler factor. Thus, this work aims to identify this scenario. Our paper is organized as follows. In section 2, we examine the connection between the Doppler factor and the synchrotron peak frequency. Section 3 presents the impact on jet power due to the Doppler factor. The Doppler-corrected blazar sequence is discussed in section 4. In section 5, we recheck the validity of examining the external Compton (EC) dominance with the correlation between CD and Doppler factor.  Discussions are presented in section 6. In this paper, we use a $\Lambda$CDM cosmology model with h=0.71, $\Omega_{m}$=0.27, $\Omega_{\Lambda}$=0.73 \citep{2009ApJS..180..330K}.

\section{Synchrotron Peak Frequency versus Doppler Factor}
The SED classification of blazars is based on the synchrotron peak frequency. The synchrotron peak frequency, $\nu_{S,p} \propto \gamma_b^2B\delta$, is related to the electron distribution and the magnetic field strength of the emission region (where $\gamma_b$ is the break energy of the electron spectrum, $B$ is the magnitude field strength, $\delta$ is the Doppler factor. See e.g., \citealt{1998ApJ...509..608T}). In order to constrain the effect on the blazar classification due to the Doppler factor, we firstly examine the correlation between the Doppler factor and the synchrotron peak frequency. Because the Doppler factor is determined by both the bulk Lorentz factor and the viewing angle ($\delta = [\Gamma (1 - \beta cos \theta)]^{-1}$, where $\beta$ is the bulk velocity in unit of the speed of light, $\Gamma$ is the bulk Lorentz factor of relativistic jet, $\theta$ is the viewing angle), the wide distribution of these two parameters for both high and low peaked sources have an intense impact on the correlation between the peak frequency and the Doppler factor. Many studies had presented that the $\gamma$-ray detected sources had smaller viewing angles (e.g. \citealt{2010A&A...512A..24S}). Thus the less aligned sources are generally excluded for a $\gamma$-ray selected sample. For a such sample, the Doppler factor is mainly determined by the intrinsic feature of jet physics, i.e. the bulk Lorentz factor. Therefore, throughout this paper, we just deal with the $\gamma$-ray selected samples, and the approximation $\delta \sim \Gamma \sim 1/\theta$ is used.

\subsection{Parameter Estimations}
The 3LAC provides an ideal $\gamma$-ray data set to cross-match with the samples of other bands. In the 3LAC, \citet{2015ApJ...810...14A} listed the synchrotron peak frequency estimated by the third-degree polynomial fit. Meanwhile, another method based on the empirical relations proposed by \citet{2010ApJ...716...30A} was often used in the literatures. \citet{2015ApJ...810...14A} compared these two methods, and concluded that the average offset between the peak frequencies estimated by these two methods was less than 0.26 dex. Because we need to estimate the bolometric luminosity and CD in the following sections, we use the empirical relations in \citet{2010ApJ...716...30A} to estimate the synchrotron peak frequency throughout this paper. The k-correction is applied as $\nu_{S,p}=\nu_{S,p}^{'}(1+z)$. The Doppler-corrected intrinsic peak frequency is calculated with $\nu_{S,int}=\nu_{S,p}/\delta$. Because the estimation of Doppler factor has large uncertainties and strong dependence on the observational epoch and frequency \citep{1999ApJ...521..493L}, we use two groups of Doppler factors estimated through two independent methods  in this paper. The two groups of results can be cross-checked. For the first method, the brightness temperature is obtained by fitting the variability timescale of radio flux \citep{1999ApJ...521..493L}, then the Doppler factor is calculated as $(T_{var}/T_{eq})^{1/3}$ (hereafter $\delta_{var}$), where $T_{eq}=5\times10^{10}K $ is the equipartition brightness temperature \citep{1994ApJ...426...51R}. For the other method, the core brightness temperature is obtained by fitting the minimum observable size \citep{1994ApJ...426...51R}, then $\delta=T_{core}/T_{eq}$ (hereafter $\delta_{eq}$).

Considering the possible variability of the Doppler factors, we search the archive for the Doppler factor during the Fermi era. $\delta_{var}$ is derived from \citet{2009A&A...494..527H}. \citet{2009A&A...494..527H} estimated $\delta_{var}$ for 89 objects (two objects were added from \citealt{2010A&A...512A..24S}). 62 of them have redshift measurement and estimation of $\nu_{S,p}$ in the 3LAC clean sample. There are 39 FSRQs, 16 LSP-BL Lacs (LBLs), 4 ISP BL Lacs (IBLs), 1 HSP-BL Lacs (HBLs) and 2 AGNs of other types in this cross-matching sample. \citet{2012ApJ...744..177L} had observed 232 AGNs with VLBA at 5GHz from 2009 to 2010, and obtained their core brightness temperature. We estimate $\delta_{eq}$ with the method described above. $\nu_{S,p}$ is obtained for 139 sources of them. Of these 139 sources, there are 85 FSRQs, 19 LBLs, 13 IBLs, 16 HBLs and 6 AGNs of other types. The details of these two samples are listed in Table 1. The two groups of Doppler factors have very different distributions (Figure 1), while $\delta_{var}$ are larger than $\delta_{eq}$ to some extent.  $\delta_{var}$ is derived from the long term radio monitoring. The targets for monitoring are biased for brighter and more variable sources \citep{2009AJ....138.1874L}, which makes HBLs and IBLs rare for that sample. Therefore, the sources with $\delta_{var}$ tend to have large Doppler factors. On the other hand, $\delta_{eq}$ estimation is only reliable for low redshift sources, and it can underestimate the real Doppler factor (see detail discussions in Section 2.2).

\subsection{Results}
The scatters of $\nu_{S,p}$ versus $\delta_{var}$ and $\nu_{S,p}$ versus $\delta_{eq}$ are plotted in Figure 1. The synchrotron peak frequency $\nu_{S,p}$ is negatively correlated with Doppler factor for both $\delta_{var}$ and $\delta_{eq}$, with the correlation coefficient of Spearman rank correlation test $\rho = - 0.48$ and the chance probability $P =  8.7\times10^{-5}$ for $\delta_{var}$, and $\rho = - 0.30$ and $P = 3.4\times10^{-4}$  for $\delta_{eq}$. In the right panel of Figure 1, there is no clear trend for each type of sources. The global correlation can be a result of different locations of HBLs and other types of sources (see follows for a possible explanation). As a result of Doppler boosting, we have $\nu_{S,p}= \delta \nu_{S,int}$. Thus the observational peak frequency is expected to  be positively correlative with Doppler factor. The opposite trends shown in Figure 1 imply some physical connections between the peak frequency and the Doppler factor.
\begin{figure*}
\begin{center}
    \includegraphics[width=7cm]{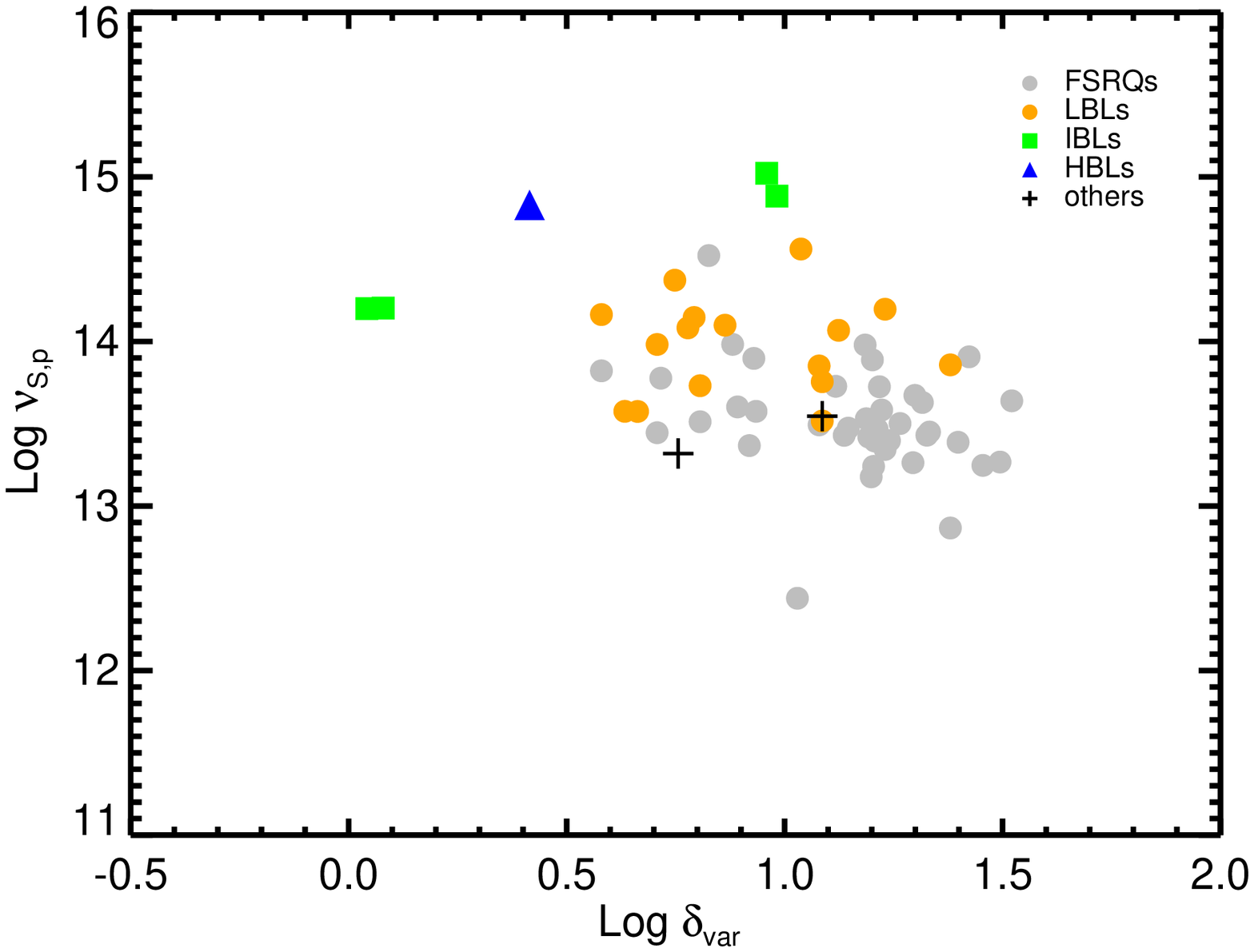}
    \includegraphics[width=7cm]{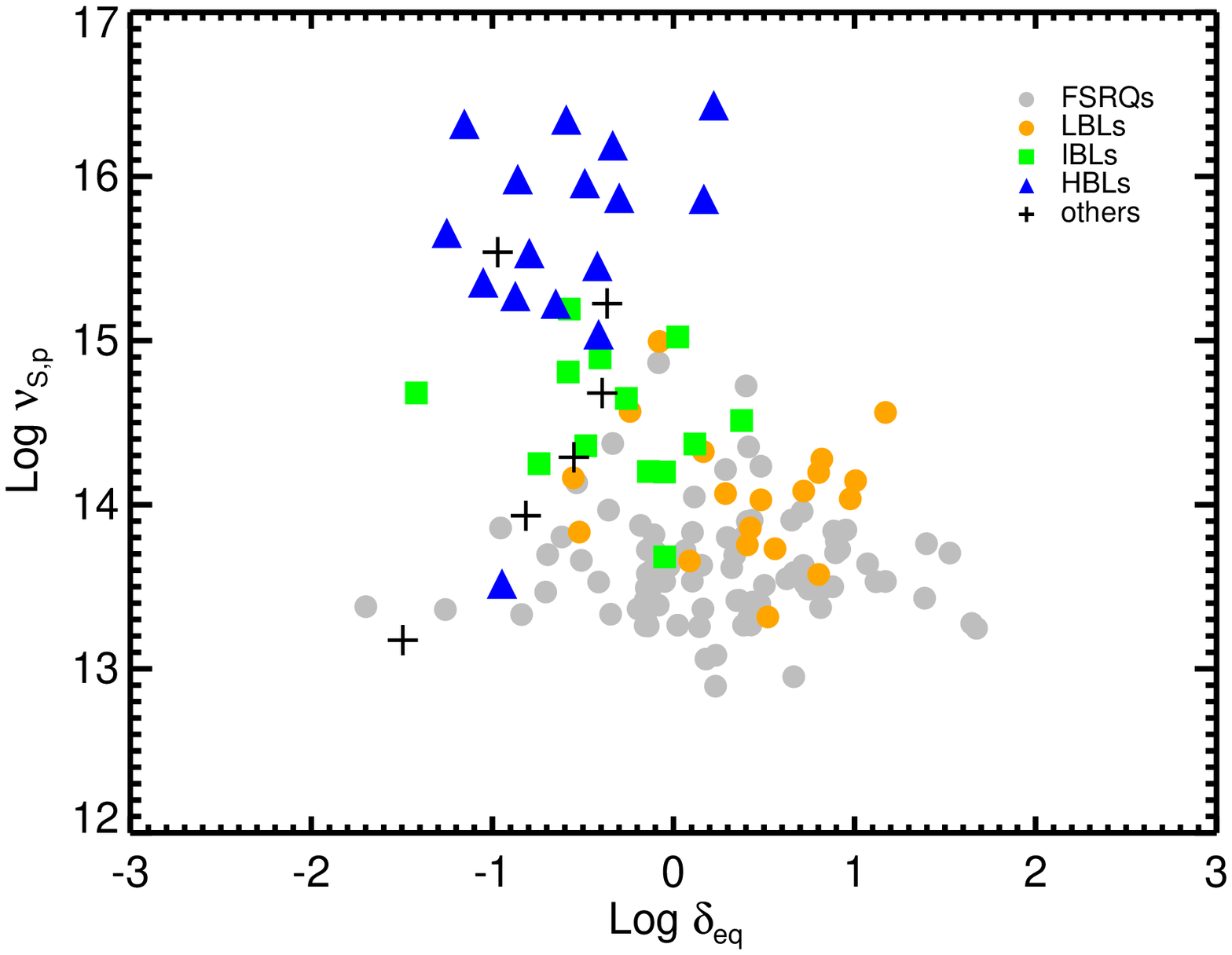}
   \caption{The correlations between the observational synchrotron peak frequency and the Doppler factor. Different classifications are represented by different symbols as labelled. Left panel: $\delta_{var}$ versus $\nu_{S,p}$. Right panel: $\delta_{eq}$ versus $\nu_{S,p}$.}
\end{center}
\end{figure*}

In order to identify the physical connections, we apply the linear regression to fit the Doppler-corrected peak frequency and the Doppler factor. The input uncertainties are set to 0.3 dex for both sides (\citealt{2015ApJ...810...14A}; \citealt{2009A&A...494..527H}). The fitting results are
\begin{equation}
log \nu_{S,int} = (-2.54\pm0.54) log \delta_{var} +(15.32\pm0.56)
\end{equation}
with the intrinsic scatter 0.14 dex, and
\begin{equation}
 log \nu_{S,int} = (-1.83\pm0.14) log \delta_{eq} +(14.09\pm0.07)
\end{equation}
with the intrinsic scatter 0.44 dex. The two linear relations are generally consistent with each other, while the latter  has a sightly flatter slope (Figure 2). \citet{1999ApJ...521..493L} proposed that during the quiescent state of total flux density, the intrinsic brightness temperature was smaller than the equipartition value. $\delta_{eq}$ is estimated through single VLBI observation, thus it underestimates the real Doppler factor for the sources passed the maximum phase of a shock development. This effect may cause the flatter slope between $\nu_{S,int}$ and $\delta_{eq}$. In the right panel of Figure 2, HBLs show different trend with other types of sources clearly. Excluding HBLs, the correlation coefficient between peak frequency and Doppler factor changes to -1.42, which is much flatter than -1.83 in Equation 2. That implies the underestimation to real Doppler factor is more serious for FSRQs and LBLs than for HBLs. One interpretation is that FSRQs and LBLs always locate at higher redshift and are more variable.
\begin{figure*}
\begin{center}
  \includegraphics[width=7cm]{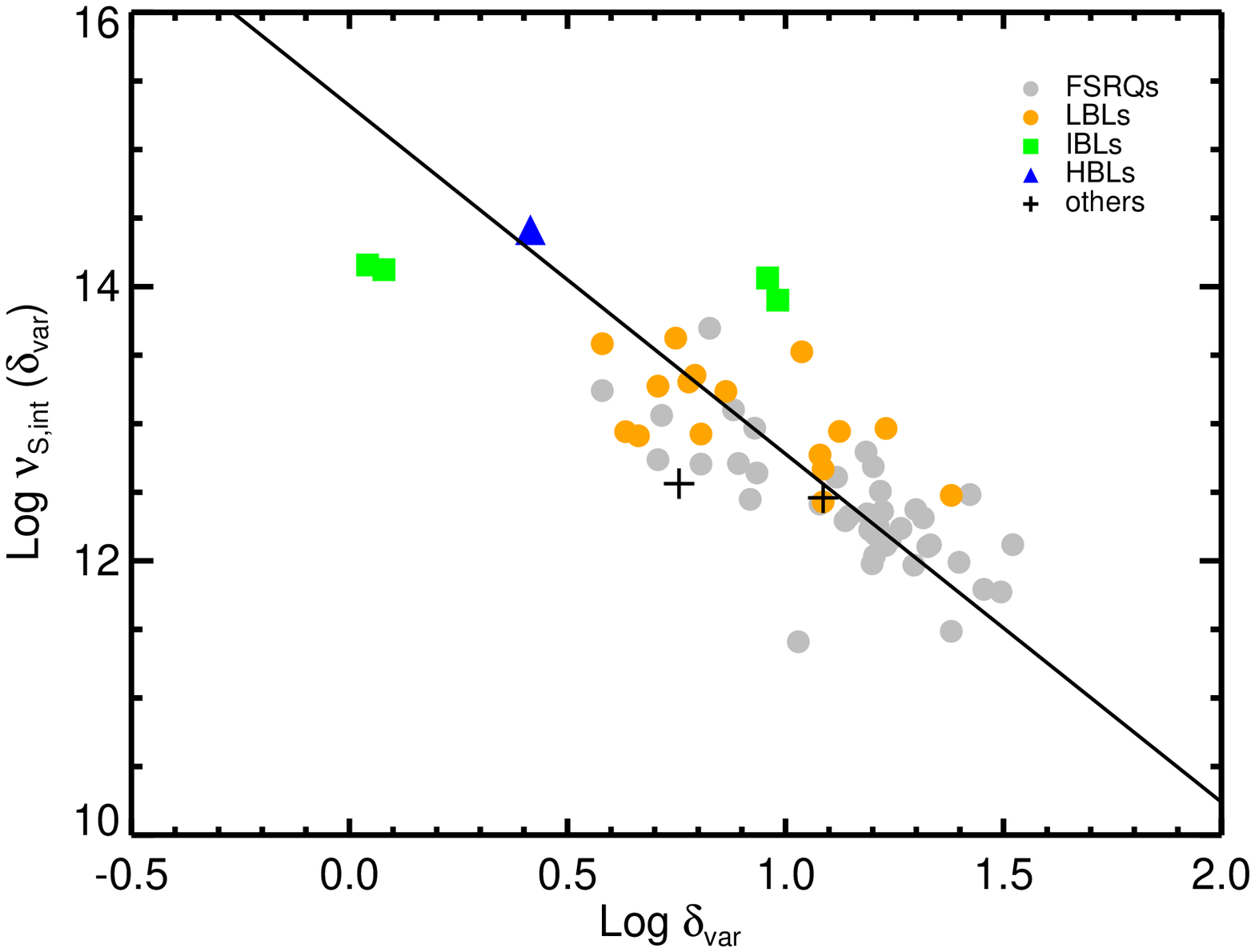}
  \includegraphics[width=7cm]{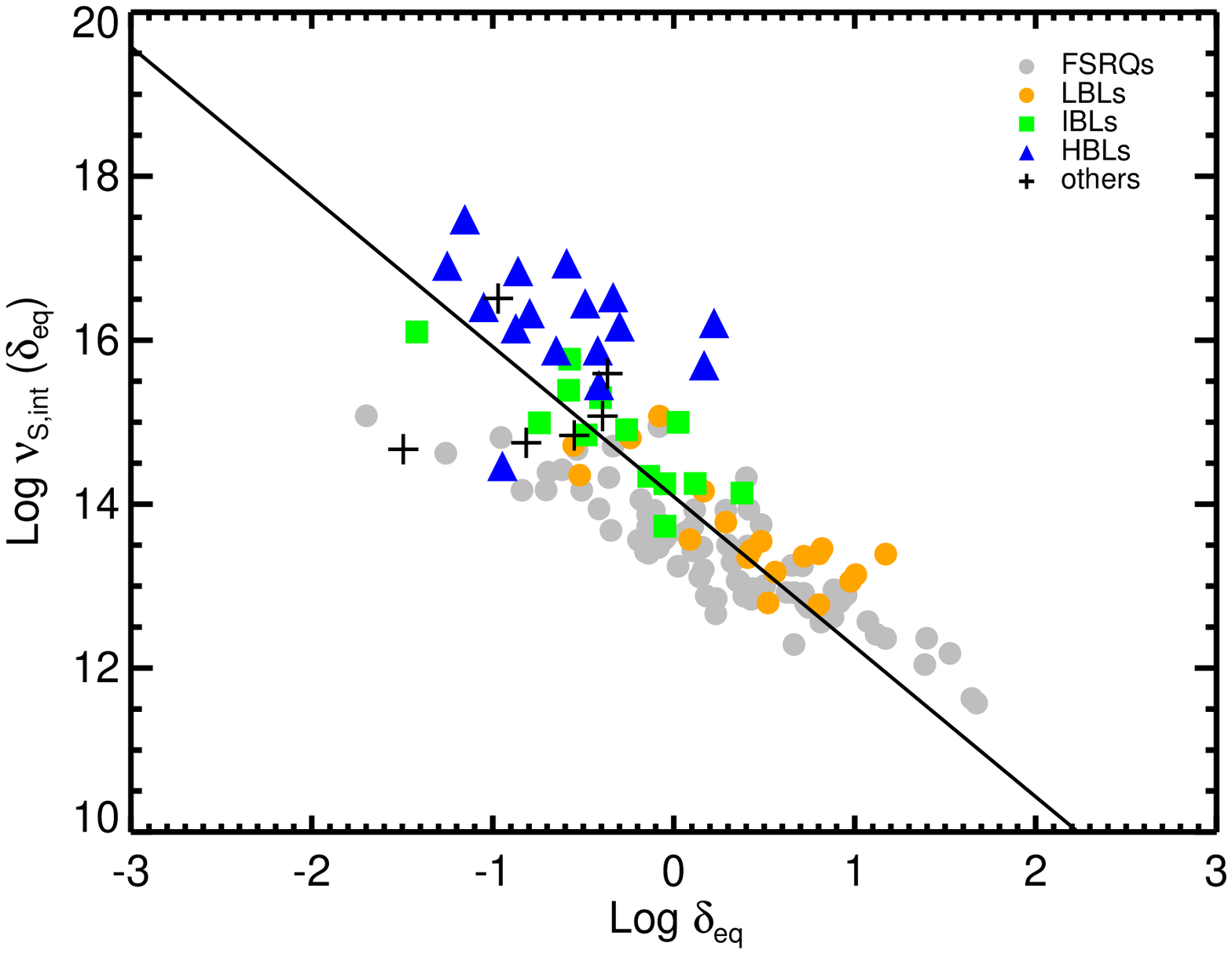}
  \caption{The correlations between the Doppler-corrected synchrotron peak frequency and the Doppler factor. Different classifications are represented by different symbols as labelled. Left panel: $\delta_{var}$ versus $\nu_{S,int}$. Right panel: $\delta_{eq}$ versus $\nu_{S,int}$. The solid lines show the best fit.}
\end{center}
\end{figure*}

\section{Jet Power}
\citet{2011ApJ...740...98M} suggested kinetic jet power as an essential feature for the blazar classifications. We thus build a cross-matching sample from the 3LAC clean sample and the sample of \citet{2012Sci...338.1445N}. This cross-matching sample contains 113 FSRQs and 104 BL Lacs. The distribution of the kinetic jet power for this sample is plotted in Figure 3. The dichotomy between FSRQs and BL Lacs is obvious shown. The Kolmogorov-Smirnov (K-S) test confirms that these two subclasses are draw from distinct samples with D = 0.67 and probability = $1.9\times10^{-22}$. Because the material energy loading in jets is also boosted by the jet speed with a factor of $\Gamma^2$ (e.g. \citealt{2008MNRAS.385..283C}), the distribution of the kinetic jet power is also influenced by the Doppler factor. We estimate the material energy in the comoving frame (hereafter intrinsic jet power) with $P_{j,int}=P_j/\Gamma^2\sim P_j/\delta^2$. The sample with $\delta_{var}$ estimation has 33 FSRQs and 20 BL Lacs, while 40 FSRQs and 44 BL Lacs have $\delta_{eq}$ estimations. The details of these two samples are also combined in Table 1. The distributions of $P_{j,int}$ calculated by two groups of $\delta$ are plotted in Figure 4. BL Lacs and FSRQs roughly have the same range of $P_{j,int}$. The null hypotheses that FSRQs and BL Lacs are draw from the same distribution are not rejected by the K - S tests, with D= 0.34, probability = 0.09 and D= 0.18, probability = 0.49 for the intrinsic jet power calculated by $\delta_{var}$ and $\delta_{eq}$, respectively. FSRQs and BL Lacs are suggested to locate in distinct accretion regimes (e.g., \citealt{2011MNRAS.414.2674G}). If this is the case, our results indicate that the material energy of jet is independent on the accretion mode. The distinctions of jet power between FSRQs and BL Lacs are mainly caused by the jet speed which is related to the jet acceleration processes or the gas environments in the host galaxies. Therefore, the assumption that the jet power is proportional to the accretion rate (e.g. \citealt{2008MNRAS.387.1669G}) should be used carefully.
\begin{figure*}
\begin{center}
    \includegraphics[width=7cm]{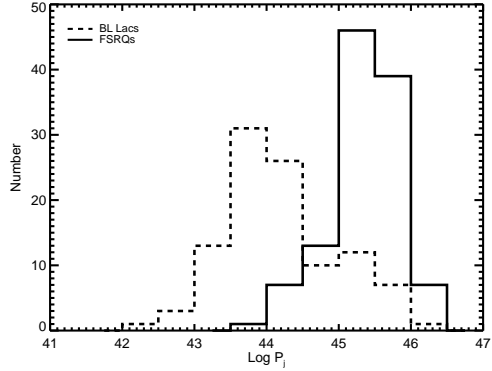}
   \caption{The distribution of the kinetic  jet power. The solid lines correspond to FSRQs, while the dashed lines correspond to BL Lacs.}
\end{center}
\end{figure*}

\begin{figure*}
\begin{center}
    \includegraphics[width=7cm]{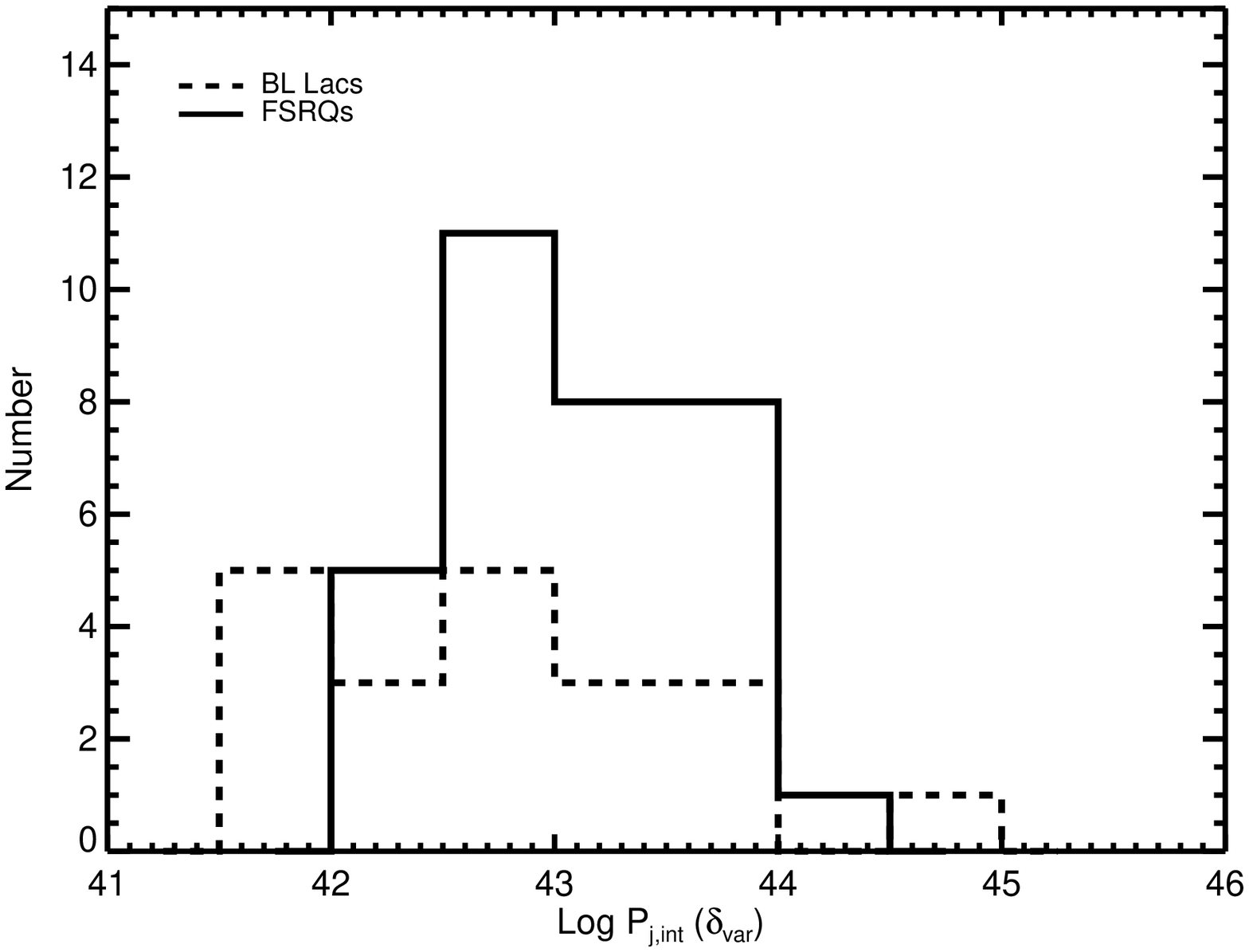}
    \includegraphics[width=7cm]{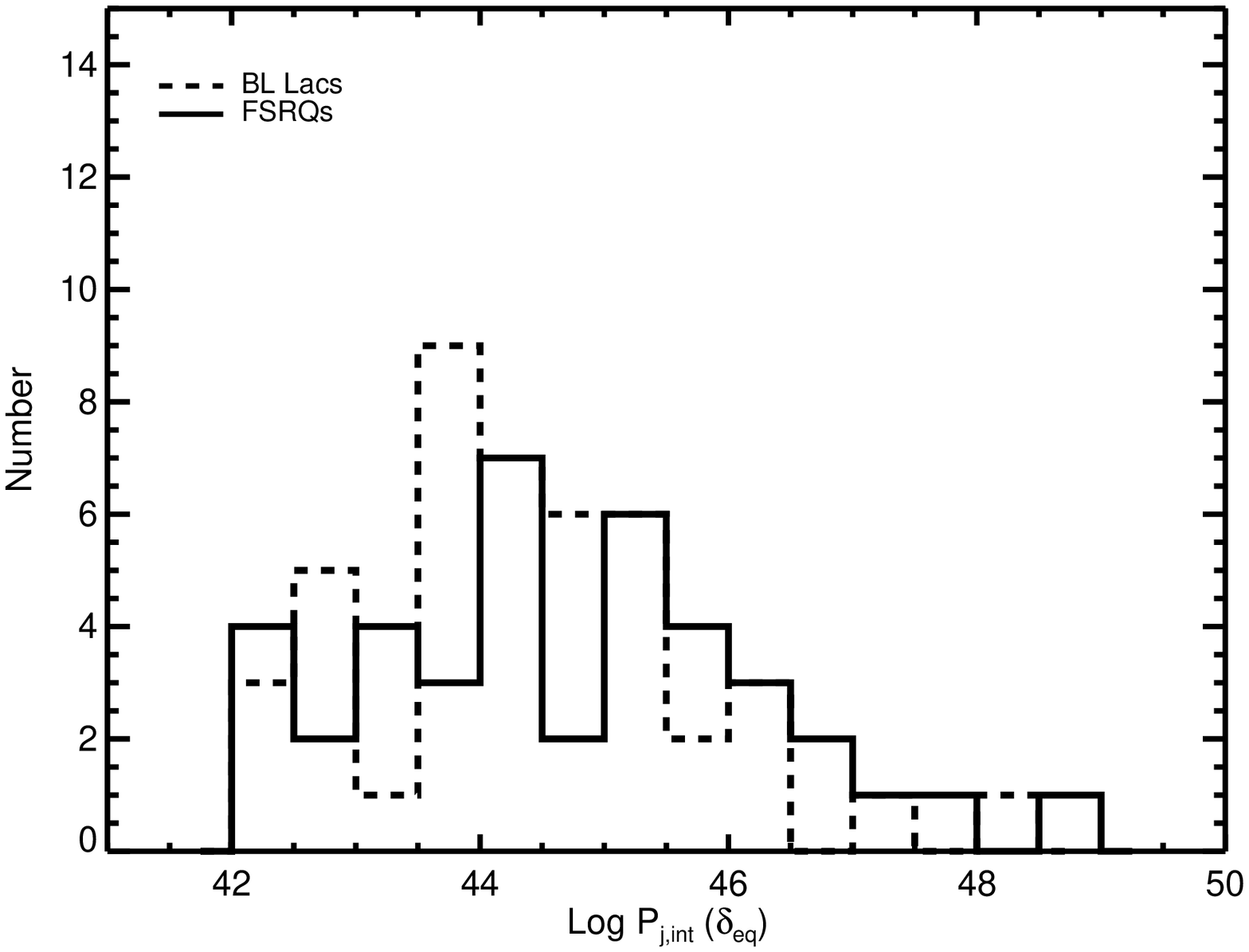}
   \caption{The distributions of the intrinsic jet power.The solid lines correspond to FSRQs, while the dashed lines correspond to BL Lacs. Left panel is calculated by $\delta_{var}$, while right panel is calculated by $\delta_{eq}$.}
\end{center}
\end{figure*}

\section{Blazar Sequence}
\citet{2008A&A...488..867N} found that the negative correlation between the peak frequency and luminosity became positive after the Doppler boosting was corrected. Thus we examine the Doppler-corrected blazar sequence for our samples. The Doppler-corrected intrinsic bolometric luminosity is calculated as follows. The isotropic luminosity $L_{iso}$ is firstly obtained by combing the synchrotron peak luminosity and IC peak luminosity, where the synchrotron peak luminosity $L_S=\nu_{S,p}L_{S,p}=4\pi{d_L}^2\nu^{'}_{S,p}F_{S,p}^{'}$, IC peak luminosity $L_{IC}=\nu_{IC,p}L_{IC,p}=4\pi{d_L}^2\nu^{'}_{IC,p}F_{IC,p}^{'}$. The peak flux of synchrotron emission $F_{S,p}^{'}$ and the peak frequency of IC process $\nu^{'}_{IC,p}$ are estimated through the empirical relations from \citet{2010ApJ...716...30A}. The flux of IC peak $F_{IC,p}^{'}$ is estimated by extrapolating the LAT flux to IC peak. Because the radiation of blazar is highly anisotropic, the realistic solid angle of the anisotropic emission is $2\pi(1-cos\theta_j)$ corresponding to the jet opening angle $2\theta_j$. However, the isotropic luminosity is calculated assuming the solid angle $4\pi$. The deviation is $(1-\cos\theta_j)/2\sim\theta_j^2/4\sim1/4\delta^2$ for small opening angles. Meanwhile, according to a moving, isotropic jet model, the boosting factor of the luminosity is considered as $\nu L_{\nu}=\delta^{4+\alpha}\nu^{iso}L^{iso}_\nu$, where $L^{iso}_\nu$ is the intrinsic monochromatic luminosity in comoving frame, $L_\nu$ is beamed monochromatic luminosity, $\alpha$ is the spectral index, which is taken as 1 around the peak (\citealt{1995PASP..107..803U}; \citealt{2008A&A...488..867N}). Thus the intrinsic bolometric luminosity is estimated as  $L_{bol}\sim L_{iso}/4\delta^7$.

\begin{figure*}
\begin{center}
    \includegraphics[width=7cm]{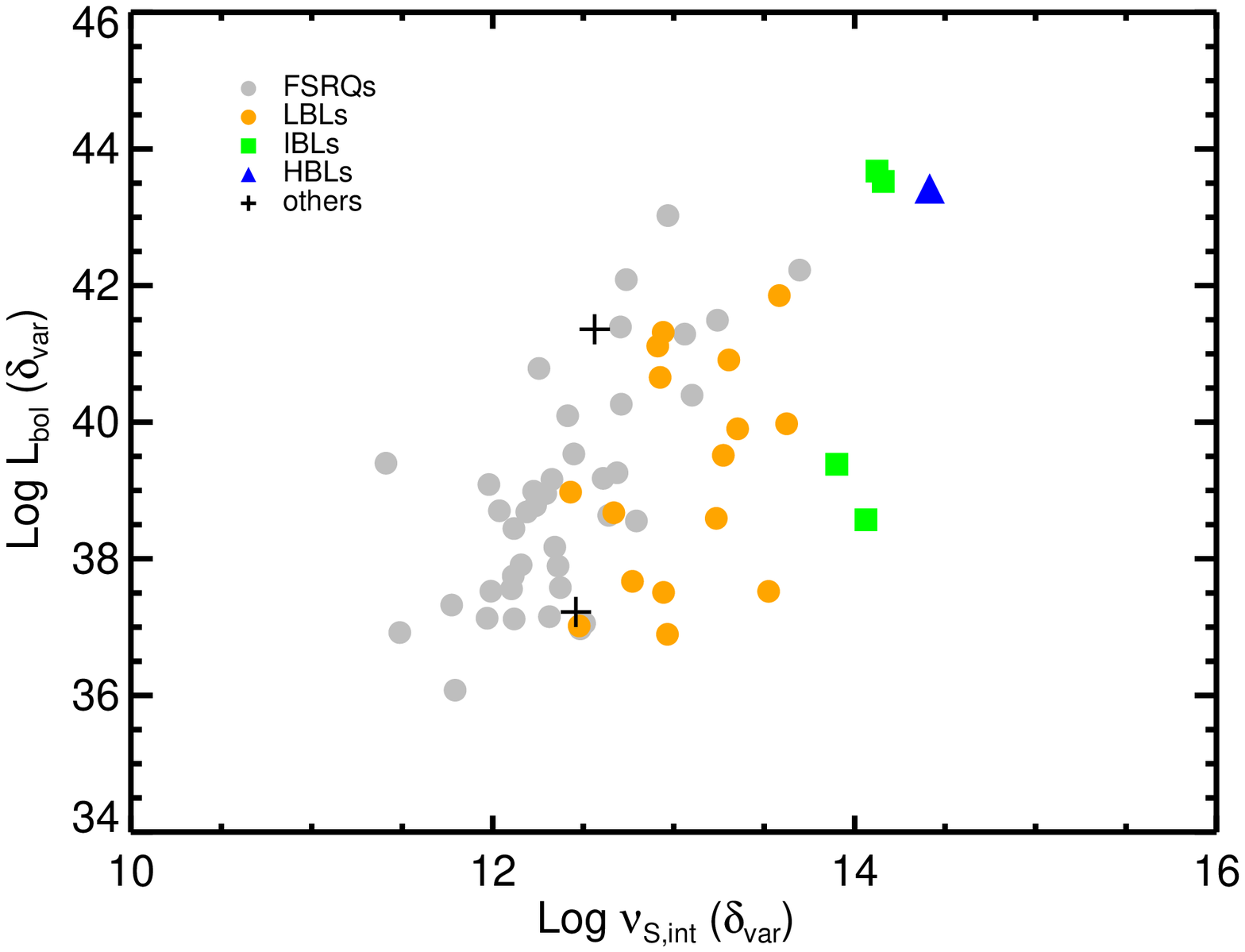}
    \includegraphics[width=7cm]{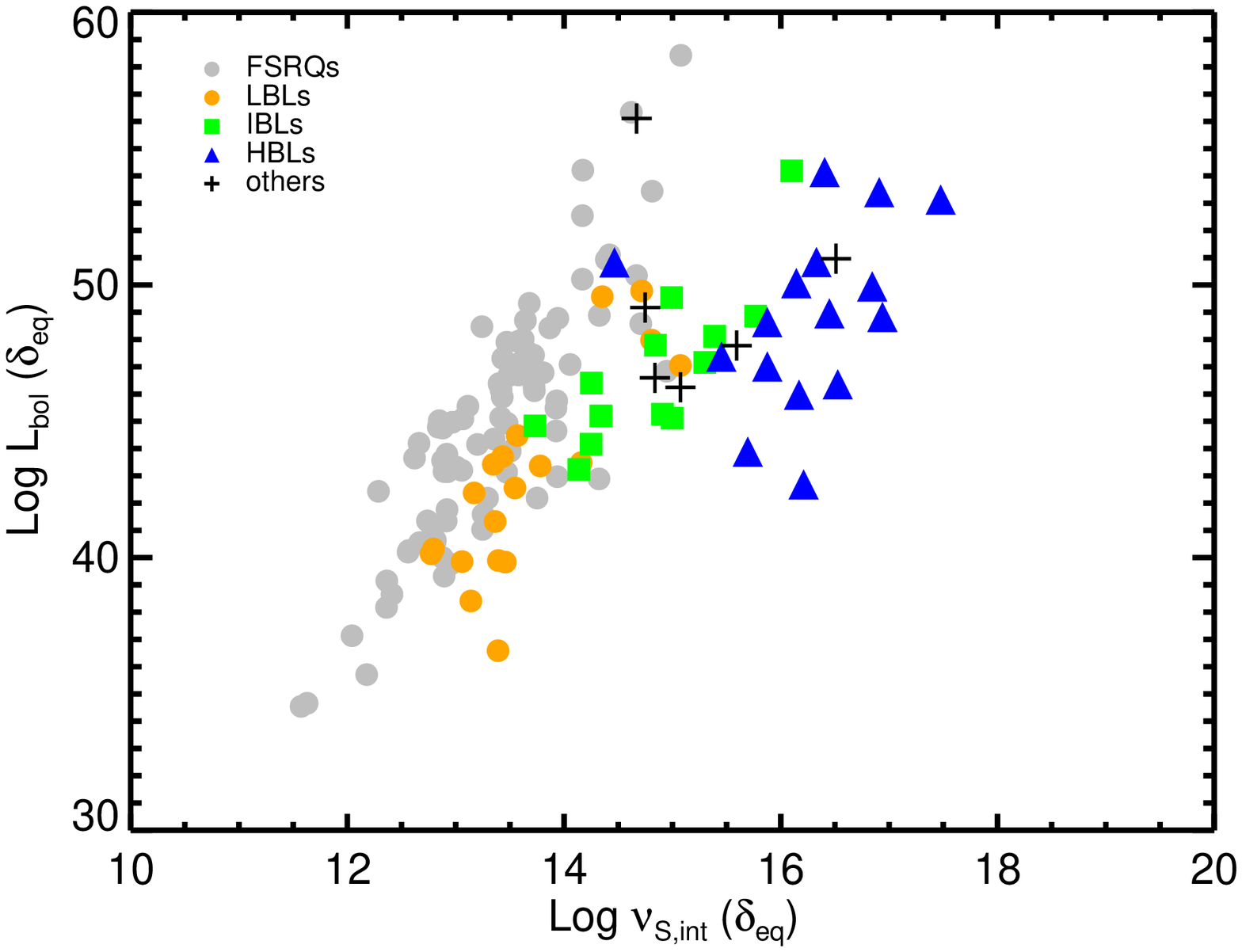}
   \caption{The correlations between the Doppler-corrected synchrotron peak frequency and bolometric luminosity. Different classifications are represented by different symbols as labelled. Left panel is calculated by $\delta_{var}$, while right panel is calculated by $\delta_{eq}$.}
\end{center}
\end{figure*}
There show apparently positive correlations between $\nu_{S,int}$ and $L_{bol}$ in Figure 5. However, after removing the common dependence on the Doppler factor of these two parameters with the partial Kendall's $\tau$ correlation test \citep{1996MNRAS.278..919A}, no correlation exists between them. The correlation coefficients $\tau$ and the chance probabilities $P$ are 0.01 and 0.87 for the parameters calculated with $\delta_{var}$, and 0.04 and 0.24 for those calculated with $\delta_{eq}$. The anti-correlation between observational peak frequency and luminosity disappears after the Doppler boosting is corrected. Thus the blazar sequence is a result of the Doppler boosting.  \citet{1998MNRAS.301..451G} explained the anti-correlation between the observational peak frequency and luminosity as the consequence of the anti-correlation between the break energy of electron spectrum and the strength of EC radiation, and proposed that the increasing cooling from the external photons led to both the decrease of the peak frequency and the increase of the luminosity. Our results indicate that the cooling effect seems unimportant to determine the peak frequency because the intrinsic luminosity has no correlation with the peak frequency. Recently, \citet{2014ApJ...788..179C} found a significant correlation between the synchrotron peak frequency and the curvature of the SEDs. Their result also implied that the break energy of electron spectrum $\gamma_b$ was mainly determined by the particle acceleration process (also see the reference therein). Then the different $\gamma_b$ result in different peak frequency. On the other hand, the different magnetic fields of the emission region also result in various observational features on peak frequency.

Similar with Figure 1 and Figure 2, HBLs show an additional track on the right panel of Figure 5. This could be caused by different physical features between HBLs and other blazars, such as the radiative efficiency or the radiation mechanism. We examine the correlation between $\nu_{S,int}$ and $L_{bol}$ for the sample excluding HBLs. The result of the partial correlation test still shows no correlation existing between these two parameters, with $\tau$ = 0.07 and $P$ = 0.07. For HBLs alone, there is also no correlation between $\nu_{S,int}$ and $L_{bol}$, with $\tau$ = 0.05 and $P$ = 0.7. On the other side, the additional track of HBLs can also be caused by the underestimation of $\delta_{eq}$ to the real Doppler factor (\citealt{1999ApJ...521..493L}, Section 2.2). Because the current sample of $\delta_{var}$ has rare HBLs, a completed sample of $\delta_{var}$ derived from the radio monitoring programmes, e.g., OVRO \citep{2011ApJS..194...29R}, would help to confirm this trend in the future.

\section{External Compton Emission}
\citet{2012ApJ...752L...4M} found a correlation between  CD and $\delta$ for the very high power sources (including radio galaxies and blazars). They explained that the high energy components of these sources were dominated by EC process rather than synchrotron self-Compton (SSC) process. Then they suggested this correlation to examine the EC dominance for blazars. We plot the scatters of CD and $\delta$ in Figure 6 for our samples. There are weak correlations between CD and $\delta$ for all sources, with $\rho$ = 0.37, $P = 3.4 \times 10^{-3}$ for $\delta_{var}$, and $\rho$ = 0.32, $P = 1.1 \times 10^{-4}$ for $\delta_{eq}$. More importantly, HBLs have the same trend on the CD-$\delta$ plane with LBLs and FSRQs (right panel of Figure 6).
\begin{figure*}
\begin{center}
    \includegraphics[width=7cm]{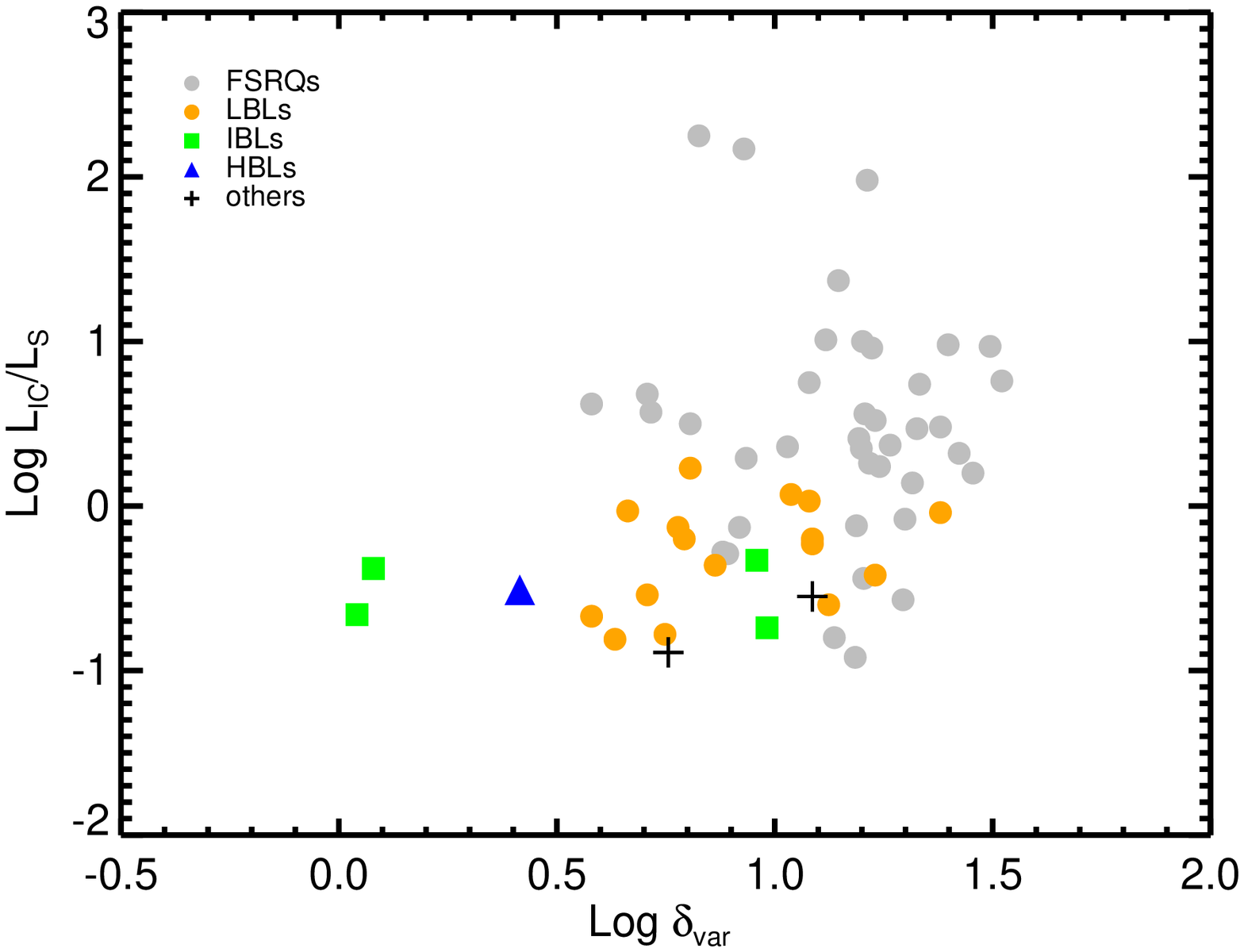}
    \includegraphics[width=7cm]{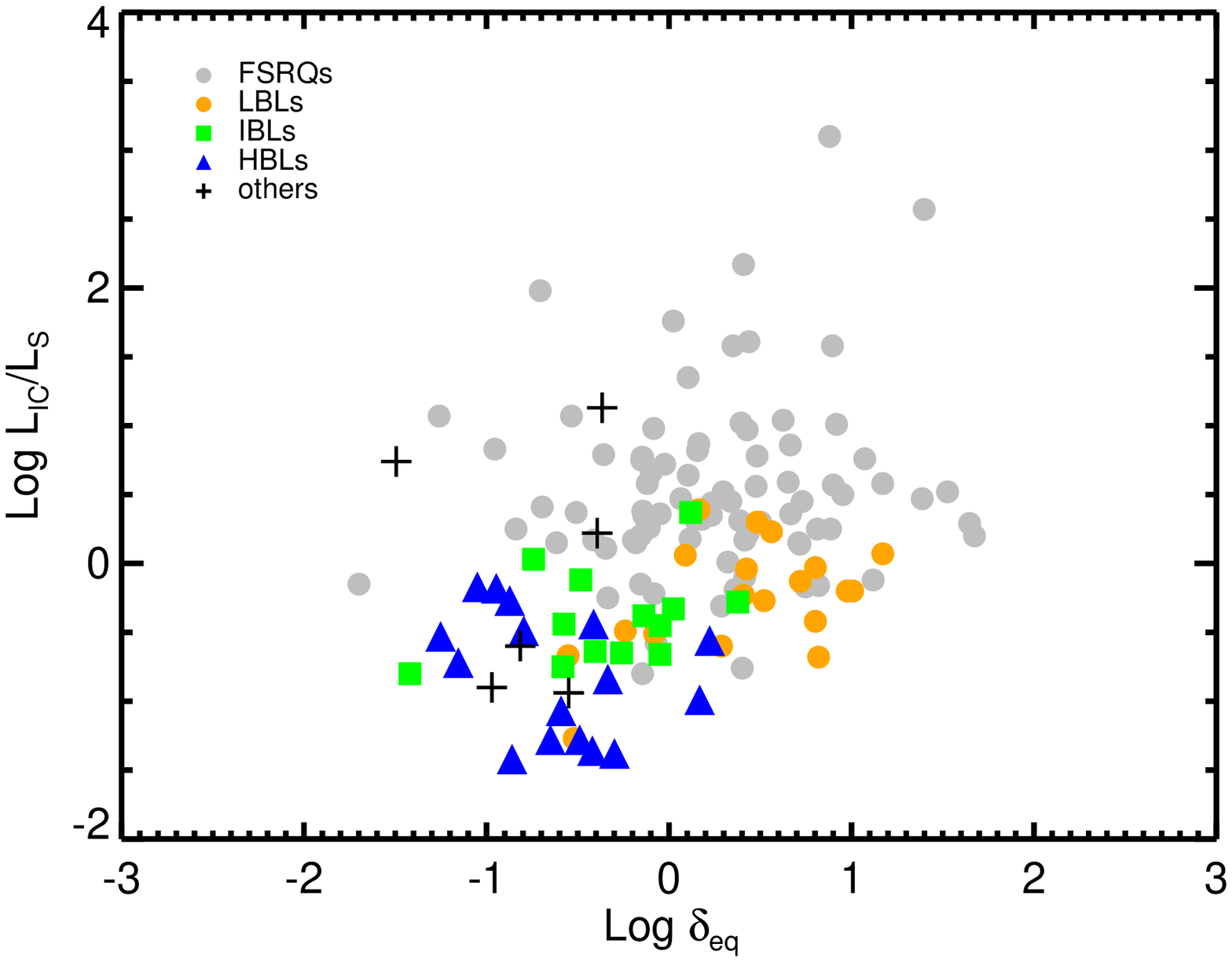}
   \caption{The correlations between the Compton dominance and the Doppler factor. Different classifications are represented by different symbols as labelled. Left panel: $\delta_{var}$ versus CD. Right panel: $\delta_{eq}$ versus CD.}
\end{center}
\end{figure*}

For SSC process, if the ratio of synchrotron energy density to magnetic field energy density $u_{syn}/u_B \simeq CD \propto \gamma_b^3N(\gamma_b)/R^2$ is independent on Doppler factor (where $N(\gamma_b)$ is the electron number at the break energy $\gamma_b$, $R$ is the radius of the emission region. See e.g. \citealt{2013ApJ...763..134F}), the correlation of CD and $\delta$ only exists for EC process. However, if either the electron distribution or the radius of emission region changes with the bulk Lorentz factor (the viewing angle seems impossible to be correlative with these parameters as a coincidence), the correlation between CD and $\delta$ is also expected for SSC process. As the result of the adiabatic expansion, the radius of emission region is the cross sectional radius of a cone.  We have $R \sim r\theta_j\sim r/\Gamma$, where $r$ is the distance between the emission region and the central nucleus, $\theta_j$ is the opening angle of the jet (e.g., \citealt{2009ApJ...704...38S}). Thus the increase of $\Gamma$ results in the decrease of $R$. As a result, CD increases as $\Gamma$ increases. This means that the correlation between CD and $\delta$ also exists for the SSC dominant sources. However, if CD correlated with $\delta$ is not unique for EC emission, this trend will be unsuitable to examine the EC dominance.

\section{Discussions}
The viewing angles of the $\gamma$-ray detected blazars are small, then the bulk Lorentz factors $\Gamma$ approximately equal to the Doppler factors $\delta$ \citep{2010A&A...512A..24S}. Thus the trends found for the Doppler factor are mainly determined by the bulk Lorentz factor. Our correlation analyses between the Doppler-corrected peak frequency and bolometric luminosity indicate that the electron acceleration processes or the magnetic field of the emission region determines the peak frequency. The correlation between the peak frequency and the Doppler factor further presents a connection between peak frequency and bulk Lorentz factor. Meanwhile, the distinction of the kinetic jet power is also caused by the bulk Lorentz factor. Therefore, we can expect that the differences of blazars are determined by a single parameter, i.e., the bulk Lorentz factor. \citet{2011ApJ...740...98M} discussed the possibility of the structured jets with velocity gradients. The radiations of the sources with large viewing angles are dominated by the slow regions. Then varying viewing angles forms an observational sequence. Similarly to that, different bulk Lorentz factors for individual blazars can directly lead to the same observational appearance.

One scenario to explain the different bulk Lorentz factors was suggested by \citet{2013MNRAS.436..304P}. They explained that the bulk Lorentz factor was governed by the accretion rate, and the black hole masses were similar for all blazars. The higher accretion rate leads to smaller mass loading into the jets, then larger fraction of the accretion power are converted to accelerate the jet. Finally it results in larger bulk Lorentz factor. Our results seems to rule out this scenario because the material energy loading in jets (i.e., the intrinsic jet power) is independent on the accretion rate (see Section 3). \citet{2012ApJ...759..114C} found a significant correlation between the bulk Lorentz factor and black hole mass, but no correlation between the bulk Lorentz factor and the Eddington ratio. Their results also suggest that the bulk Lorentz factor is independent on the accretion rate, but governed by the black hole mass. \citet{2013MNRAS.431.1840P} also presented another scenario to unify the jet physics. They assumed that the radius of transition region (where the jet comes into equipartition between magnetic filed energy and particle energy, and dominates the optically thin synchrotron emission) scaled linearly with black hole mass. Therefore, the larger black hole mass leads to the farer transition region from the central black hole. This further results in lower magnetic field strength, finally results in lower synchrotron peak frequency. Based on another assumption that the intrinsic jet power has a fixed fraction of the Eddington luminosity (i.e. the intrinsic jet power is independent on accretion rate), they got a relation between the synchrotron peak frequency and the black hole mass with the form $\nu_{S,int} \propto M_{BH}^{-1/2}$. Observationally, The anti-correlation between the peak frequency and black hole mass has been found by e.g., \citet{2011ApJ...735..108C}. When combined this relation with $\Gamma \propto M_{BH}^{0.2}$ found by \citet{2012ApJ...759..114C},  we have $\nu_{S,int} \propto \Gamma^{-2.5}$. This relation agrees well with our findings presented in Equations (1) and (2). Therefore, the SEDs of blazars are mainly determined by the black hole mass, but not the luminosity \citep{1998MNRAS.301..451G}, accretion rate \citep{2009MNRAS.396L.105G}, or orientation \citep{2011ApJ...740...98M}.

\section{Conclusion}
In this paper, we obtain two groups of Doppler factors estimated through two independent methods, aim to identify whether the Doppler factor determines the observational differences of blazars. Significant correlations are found between the Doppler factors and the indicator of the SED classification, i.e., observational synchrotron peak frequency. After corrected the Doppler boosting, the intrinsic peak frequency has uniform linear relations with two groups of Doppler factors. In addition, we find the distinction of jet power is mainly caused by the different Doppler factors for different subclasses. The negative correlation between the peak frequency and the observational isotropic luminosity disappears after the Doppler boosting is corrected. All these results confirm that the Doppler factor (physically the bulk Lorentz factor) determines the observational differences of blazars. Furthermore, the black hole mass plays an important role to control the bulk Lorentz factor and SED of blazars. Moreover, we find the correlation between the Compton dominance and the Doppler factor existing for all types of blazars, thus this correlation is unsuitable to examine the EC emission dominance. The correlation between CD and Doppler factor can be explained for SSC process if the radius of emission region decreases as the bulk Lorentz factor increases.

\begin{acknowledgements}
We thank the anonymous referee for his/her useful comments. We are also grateful to Yi-Bo Wang, Liang Chen and Neng-Hui Liao for their useful discussions which improve this manuscript intensively. This research is supported by the Strategic Priority Research Program of the Chinese Academy of Sciences - The Emergence of Cosmological Structures (grant No. XDB09000000), the Key Research Program of the Chinese Academy of Sciences (grant No. KJZD-EW-M06), and the NSFC through NSFC-11133006 and 11361140347. J. Mao is supported by the Hundred-Talent Program of Chinese Academy of Sciences.
\end{acknowledgements}

\clearpage

\setcounter{table}{0}
\begin{table}
\bc
\caption{Source Data}
\setlength{\tabcolsep}{3pt}
\small
 \begin{tabular}{cccccccccccccc}
  \hline\noalign{\smallskip}
3FGL name  & z    & opt.     &  SED    &
 $\nu_{S,p}$    &  $CD$    &  $\delta_{var}$   &
 $\nu_{S,int}$     &  $L_{bol}$      &  $P_{j,int}$   &
 $\delta_{eq}$   &  $\nu_{S,int}$     &  $L_{bol}$    &  $P_{j,int}$ \\
  \hline\noalign{\smallskip}
J0006.4+3825 & 0.23 &   FSRQ & LSP & 13.58 &  0.77 &  - &  - &  - &  - & -0.14 & 13.72 & 46.25 &  - \\
J0050.6-0929 & 0.63 & BL Lac & ISP & 14.88 & -0.74 &  0.98 & 13.90 & 39.38 & 42.35 &  - &  - &  - &  - \\
J0058.3+3315 & 1.37 &   FSRQ & LSP & 13.51 &  0.45 &  - &  - &  - &  - &  0.73 & 12.79 & 41.10 &  - \\
J0108.7+0134 & 2.10 &   FSRQ & LSP & 13.50 &  0.37 &  1.26 & 12.24 & 38.78 & 43.93 &  - &  - &  - &  - \\
J0109.1+1816 & 0.44 & BL Lac & HSP & 15.96 & -1.28 &  - &  - &  - &  - & -0.49 & 16.45 & 48.96 &  - \\
J0112.1+2245 & 0.26 & BL Lac & ISP & 15.02 & -0.33 &  0.96 & 14.06 & 38.57 & 41.83 &  0.02 & 15.00 & 45.11 & 43.70 \\
J0112.8+3207 & 0.60 &   FSRQ & ISP & 14.86 & -0.22 &  - &  - &  - &  - & -0.08 & 14.95 & 46.84 & 45.37 \\
J0113.4+4948 & 0.39 &   FSRQ & LSP & 13.26 & -0.15 &  - &  - &  - &  - & -0.16 & 13.42 & 46.14 &  - \\
J0137.0+4752 & 0.86 &   FSRQ & LSP & 13.63 &  0.14 &  1.32 & 12.31 & 37.15 & 42.15 &  0.72 & 12.91 & 41.34 & 43.34 \\
J0151.6+2205 & 1.32 &   FSRQ & LSP & 13.45 &  0.68 &  0.71 & 12.74 & 42.09 &  - &  - &  - &  - &  - \\
J0204.8+3212 & 1.47 &   FSRQ & LSP & 13.76 &  2.57 &  - &  - &  - &  - &  1.40 & 12.36 & 39.14 & 42.53 \\
J0217.5+7349 & 2.37 &   FSRQ & LSP & 13.90 &  2.17 &  0.93 & 12.97 & 43.02 & 43.15 &  0.41 & 13.49 & 46.67 & 44.19 \\
J0221.1+3556 & 0.94 &   FSRQ & LSP &  - &  - &  - &  - &  - &  - & -1.17 &  - &  - & 47.91 \\
J0222.6+4301 & 0.44 & BL Lac & HSP & 14.83 & -0.51 &  0.41 & 14.41 & 43.42 & 44.60 &  - &  - &  - &  - \\
J0237.9+2848 & 1.21 &   FSRQ & LSP & 13.39 &  0.56 &  1.21 & 12.19 & 38.69 & 43.18 &  0.48 & 12.92 & 43.79 & 44.64 \\
J0238.6+1636 & 0.94 & BL Lac & LSP & 13.86 & -0.04 &  1.38 & 12.48 & 37.02 & 41.96 &  0.42 & 13.43 & 43.71 & 43.87 \\
J0245.4+2410 & 2.24 &   FSRQ & LSP & 13.36 &  1.07 &  - &  - &  - &  - & -1.26 & 14.62 & 56.33 &  - \\
J0310.8+3814 & 0.94 &   FSRQ & LSP & 13.96 &  0.15 &  - &  - &  - &  - &  0.71 & 13.25 & 41.03 &  - \\
J0325.2+3410 & 0.06 &  NLSY1 & HSP & 15.23 &  1.13 &  - &  - &  - &  - & -0.37 & 15.59 & 47.77 &  - \\
J0325.5+2223 & 2.07 &   FSRQ & LSP & 13.41 &  1.61 &  - &  - &  - &  - &  0.44 & 12.97 & 44.96 &  - \\
J0339.5-0146 & 0.85 &   FSRQ & LSP & 13.40 &  0.24 &  1.24 & 12.16 & 37.91 & 42.83 &  - &  - &  - &  - \\
J0423.2-0119 & 0.92 &   FSRQ & LSP & 13.67 & -0.08 &  1.30 & 12.37 & 37.58 & 42.64 &  - &  - &  - &  - \\
J0433.6+2905 & 0.97 & BL Lac & LSP & 14.99 & -0.51 &  - &  - &  - &  - & -0.08 & 15.07 & 47.05 &  - \\
J0449.0+1121 & 2.15 &   FSRQ & LSP & 13.81 &  1.02 &  - &  - &  - &  - &  0.39 & 13.41 & 45.14 & 44.43 \\
J0501.2-0157 & 2.29 &   FSRQ & LSP & 13.18 &  0.35 &  1.20 & 11.98 & 39.09 & 43.90 &  - &  - &  - &  - \\
J0509.3+1012 & 0.62 &   FSRQ & LSP & 13.71 &  0.66 &  - &  - &  - &  - & -0.10 & 13.81 & 46.78 &  - \\
J0510.0+1802 & 0.42 &   FSRQ & LSP & 13.26 &  0.35 &  - &  - &  - &  - & -0.14 & 13.40 & 46.38 &  - \\
J0530.8+1330 & 2.07 &   FSRQ & LSP & 13.27 &  0.97 &  1.49 & 11.77 & 37.32 & 42.93 &  0.43 & 12.84 & 44.79 & 45.06 \\
J0608.0-0835 & 0.87 &   FSRQ & ISP & 13.98 & -0.28 &  0.88 & 13.10 & 40.40 & 43.66 &  - &  - &  - &  - \\
J0612.8+4122 & - & BL Lac & ISP &  - &  - &  - &  - &  - &  - &  0.02 &  - &  - & 43.70 \\
J0638.6+7324 & 1.85 &   FSRQ & LSP & 13.62 &  0.72 &  - &  - &  - &  - & -0.02 & 13.65 & 47.59 &  - \\
J0650.7+2503 & 0.20 & BL Lac & HSP & 16.32 & -0.72 &  - &  - &  - &  - & -1.15 & 17.48 & 53.13 &  - \\
J0654.4+4514 & 0.93 &   FSRQ & LSP &  - &  - &  - &  - &  - &  - & -0.48 &  - &  - & 46.05 \\
J0654.4+5042 & 1.25 &   FSRQ & ISP & 14.37 & -0.25 &  - &  - &  - &  - & -0.34 & 14.71 & 48.58 & 44.29 \\
J0710.5+4732 & 1.29 & BL Lac & ISP & 14.37 &  0.37 &  - &  - &  - &  - &  0.12 & 14.25 & 46.40 & 45.35 \\
J0712.6+5033 & 0.50 & BL Lac & LSP & 13.66 &  0.06 &  - &  - &  - &  - &  0.09 & 13.57 & 44.48 &  - \\
J0719.3+3307 & 0.78 &   FSRQ & ISP & 13.80 &  0.52 &  - &  - &  - &  - &  0.30 & 13.50 & 43.90 &  - \\
J0721.9+7120 & 0.13 & BL Lac & LSP & 14.56 &  0.07 &  1.04 & 13.52 & 37.52 & 42.69 &  1.17 & 13.39 & 36.59 & 42.42 \\
J0725.2+1425 & 1.04 &   FSRQ & ISP & 13.60 &  0.28 &  - &  - &  - &  - & -0.12 & 13.72 & 47.43 &  - \\
J0738.1+1741 & 0.42 & BL Lac & LSP & 14.16 & -0.67 &  0.58 & 13.58 & 41.85 & 43.31 & -0.55 & 14.72 & 49.78 & 45.58 \\
J0739.4+0137 & 0.19 &   FSRQ & ISP & 13.58 &  0.29 &  0.93 & 12.64 & 38.64 & 42.23 &  - &  - &  - &  - \\
J0742.6+5444 & 0.72 &   FSRQ & ISP & 13.85 &  0.50 &  - &  - &  - &  - &  0.95 & 12.89 & 39.32 &  - \\
J0746.4+2540 & 2.98 &   FSRQ & LSP & 13.50 &  3.10 &  - &  - &  - &  - &  0.88 & 12.62 & 43.65 &  - \\
  \noalign{\smallskip}\hline
\end{tabular}
\ec
\end{table}

\setcounter{table}{0}
\begin{table}
\bc
\caption{continued.}
\setlength{\tabcolsep}{3pt}
\small
 \begin{tabular}{cccccccccccccc}
  \hline\noalign{\smallskip}
3FGL name  & z    & opt.     &  SED    &
 $\nu_{S,p}$    &  $CD$    &  $\delta_{var}$   &
 $\nu_{S,int}$     &  $L_{bol}$      &  $P_{j,int}$   &
 $\delta_{eq}$   &  $\nu_{S,int}$     &  $L_{bol}$    &  $P_{j,int}$ \\
  \hline\noalign{\smallskip}
J0750.6+1232 & 0.89 &   FSRQ & LSP & 13.49 & -0.16 &  - &  - &  - &  - &  0.82 & 12.67 & 40.55 &  - \\
J0757.0+0956 & 0.27 & BL Lac & LSP & 14.37 & -0.78 &  0.75 & 13.62 & 39.98 & 42.84 &  - &  - &  - &  - \\
J0805.4+6144 & 3.03 &   FSRQ & LSP & 13.27 &  1.76 &  - &  - &  - &  - &  0.02 & 13.24 & 48.47 &  - \\
J0809.6+3456 & 0.08 & BL Lac & HSP & 15.45 & -1.36 &  - &  - &  - &  - & -0.42 & 15.87 & 47.01 &  - \\
J0809.8+5218 & 0.14 & BL Lac & HSP & 15.66 & -0.53 &  - &  - &  - &  - & -1.25 & 16.91 & 53.40 & 45.79 \\
J0814.7+6428 & 0.24 & BL Lac & LSP & 14.32 &  0.39 &  - &  - &  - &  - &  0.16 & 14.16 & 43.48 &  - \\
J0816.7+5739 & - & BL Lac & HSP &  - &  - &  - &  - &  - &  - & -2.00 &  - &  - & 48.01 \\
J0818.2+4223 & 0.53 & BL Lac & LSP & 13.57 & -0.03 &  0.66 & 12.91 & 41.11 & 43.08 &  0.80 & 12.77 & 40.14 & 42.81 \\
J0824.9+5551 & 1.42 &   FSRQ & LSP & 13.50 &  0.58 &  - &  - &  - &  - & -0.12 & 13.62 & 48.01 &  - \\
J0830.7+2408 & 0.94 &   FSRQ & LSP & 13.73 &  1.01 &  1.12 & 12.61 & 39.18 & 42.99 &  0.92 & 12.81 & 40.57 & 43.38 \\
J0834.1+4223 & 0.25 &   FSRQ & ISP & 13.82 &  0.26 &  - &  - &  - &  - & -0.11 & 13.92 & 45.48 & 44.40 \\
J0841.4+7053 & 2.22 &   FSRQ & LSP & 13.47 &  1.98 &  1.21 & 12.26 & 40.79 & 43.55 & -0.71 & 14.17 & 54.21 & 47.38 \\
J0850.2-1214 & 0.57 &   FSRQ & LSP & 13.72 &  0.26 &  1.22 & 12.51 & 37.06 &  - &  - &  - &  - &  - \\
J0854.8+2006 & 0.31 & BL Lac & LSP & 14.20 & -0.42 &  1.23 & 12.97 & 36.89 & 41.71 &  0.80 & 13.39 & 39.89 & 42.57 \\
J0915.8+2933 & - & BL Lac & HSP &  - &  - &  - &  - &  - &  - & -0.59 &  - &  - & 44.93 \\
J0920.9+4442 & 2.19 &   FSRQ & LSP & 13.72 &  0.38 &  - &  - &  - &  - & -0.14 & 13.87 & 48.43 & 46.13 \\
J0921.8+6215 & 1.45 &   FSRQ & LSP & 13.51 &  0.30 &  - &  - &  - &  - &  0.50 & 13.01 & 43.32 & 44.66 \\
J0929.4+5013 & - & BL Lac & ISP &  - &  - &  - &  - &  - &  - & -0.02 &  - &  - & 44.13 \\
J0937.7+5008 & 0.28 &   FSRQ & LSP & 13.62 &  0.01 &  - &  - &  - &  - &  0.32 & 13.29 & 42.18 &  - \\
J0945.9+5756 & 0.23 & BL Lac & ISP & 14.36 & -0.12 &  - &  - &  - &  - & -0.48 & 14.84 & 47.79 & 44.79 \\
J0948.6+4041 & 1.25 &   FSRQ & LSP & 13.51 &  0.50 &  0.81 & 12.71 & 41.39 & 44.04 &  - &  - &  - &  - \\
J0957.6+5523 & 0.90 &   FSRQ & ISP & 13.38 & -0.15 &  - &  - &  - &  - & -1.70 & 15.08 & 58.42 & 48.96 \\
J0958.6+6534 & 0.37 & BL Lac & LSP & 14.15 & -0.20 &  0.79 & 13.35 & 39.90 & 42.84 &  1.01 & 13.14 & 38.41 & 42.41 \\
J1012.6+2439 & 1.80 &   FSRQ & LSP & 13.86 &  0.83 &  - &  - &  - &  - & -0.95 & 14.81 & 53.44 &  - \\
J1015.0+4925 & 0.21 & BL Lac & HSP & 15.53 & -0.49 &  - &  - &  - &  - & -0.80 & 16.33 & 50.84 &  - \\
J1033.2+4116 & 1.12 &   FSRQ & LSP & 13.28 &  0.29 &  - &  - &  - &  - &  1.65 & 11.63 & 34.66 & 42.06 \\
J1033.8+6051 & 1.40 &   FSRQ & LSP & 13.66 &  0.37 &  - &  - &  - &  - & -0.51 & 14.17 & 50.21 & 46.53 \\
J1037.5+5711 & - & BL Lac & ISP &  - &  - &  - &  - &  - &  - & -0.46 &  - &  - & 44.26 \\
J1043.1+2407 & 0.56 &   FSRQ & LSP & 14.72 & -0.76 &  - &  - &  - &  - &  0.40 & 14.32 & 42.89 &  - \\
J1048.4+7144 & 1.15 &   FSRQ & LSP & 13.69 &  0.45 &  - &  - &  - &  - &  0.34 & 13.35 & 44.35 &  - \\
J1058.5+0133 & 0.89 & BL Lac & LSP & 13.52 & -0.20 &  1.09 & 12.43 & 38.98 & 43.44 &  - &  - &  - &  - \\
J1058.6+5627 & 0.14 & BL Lac & HSP & 15.04 & -0.44 &  - &  - &  - &  - & -0.41 & 15.45 & 47.37 & 44.51 \\
J1104.4+3812 & 0.03 & BL Lac & HSP & 16.43 & -0.56 &  - &  - &  - &  - &  0.22 & 16.21 & 42.68 & 42.83 \\
J1105.9+2814 & 0.84 &   FSRQ & LSP & 13.91 &  0.59 &  - &  - &  - &  - &  0.65 & 13.25 & 41.58 &  - \\
J1112.4+3449 & 1.96 &   FSRQ & LSP & 13.83 &  0.64 &  - &  - &  - &  - &  0.11 & 13.73 & 46.12 &  - \\
J1117.0+2014 & 0.14 & BL Lac & HSP & 16.35 & -1.07 &  - &  - &  - &  - & -0.59 & 16.94 & 48.83 &  - \\
J1124.1+2337 & 1.55 &   FSRQ & LSP & 13.26 &  0.32 &  - &  - &  - &  - &  0.14 & 13.11 & 45.55 &  - \\
J1136.6+7009 & 0.05 & BL Lac & HSP & 15.87 & -1.38 &  - &  - &  - &  - & -0.30 & 16.17 & 45.97 & 44.08 \\
J1150.3+2417 & 0.18 & BL Lac & ISP & 13.68 & -0.45 &  - &  - &  - &  - & -0.05 & 13.73 & 44.83 &  - \\
J1151.4+5858 & - & BL Lac & ISP &  - &  - &  - &  - &  - &  - & -1.11 &  - &  - & 46.12 \\
J1154.3+6023 & 1.12 &   FSRQ & LSP & 13.71 &  1.58 &  - &  - &  - &  - &  0.90 & 12.81 & 40.65 & 42.72 \\
J1159.5+2914 & 0.73 &   FSRQ & LSP & 13.25 &  0.20 &  1.45 & 11.79 & 36.08 & 42.52 &  1.67 & 11.57 & 34.54 & 42.08 \\
  \noalign{\smallskip}\hline
\end{tabular}
\ec
\end{table}

\setcounter{table}{0}
\begin{table}
\bc
\caption{continued.}
\setlength{\tabcolsep}{3pt}
\small
 \begin{tabular}{cccccccccccccc}
  \hline\noalign{\smallskip}
3FGL name  & z    & opt.     &  SED    &
 $\nu_{S,p}$    &  $CD$    &  $\delta_{var}$   &
 $\nu_{S,int}$     &  $L_{bol}$      &  $P_{j,int}$   &
 $\delta_{eq}$   &  $\nu_{S,int}$     &  $L_{bol}$    &  $P_{j,int}$ \\
  \hline\noalign{\smallskip}
J1203.1+6029 & 0.07 & BL Lac & ISP & 14.65 & -0.65 &  - &  - &  - &  - & -0.26 & 14.91 & 45.26 & 43.97 \\
J1209.4+4119 & - & BL Lac & LSP &  - &  - &  - &  - &  - &  - & -0.24 &  - &  - & 43.62 \\
J1217.8+3007 & 0.13 & BL Lac & HSP & 15.86 & -0.99 &  - &  - &  - &  - &  0.17 & 15.70 & 43.88 & 43.58 \\
J1220.2+7105 & 0.45 &   FSRQ & ISP &  - &  - &  - &  - &  - &  - &  0.30 &  - &  - & 44.06 \\
J1221.4+2814 & 0.10 & BL Lac & ISP & 14.20 & -0.38 &  0.08 & 14.12 & 43.68 & 41.98 & -0.14 & 14.34 & 45.19 & 42.42 \\
J1224.6+4332 & - & BL Lac & LSP &  - &  - &  - &  - &  - &  - & -0.20 &  - &  - & 46.14 \\
J1224.9+2122 & 0.44 &   FSRQ & LSP & 13.78 &  0.57 &  0.72 & 13.06 & 41.29 & 43.95 &  0.90 & 12.88 & 40.00 & 43.58 \\
J1229.1+0202 & 0.16 &   FSRQ & LSP & 13.34 &  0.52 &  1.23 & 12.11 & 37.75 & 43.04 &  - &  - &  - &  - \\
J1230.3+2519 & 0.14 & BL Lac & ISP & 14.89 & -0.64 &  - &  - &  - &  - & -0.40 & 15.30 & 47.16 &  - \\
J1231.7+2847 & 0.24 & BL Lac & ISP & 15.19 & -0.44 &  - &  - &  - &  - & -0.58 & 15.77 & 48.86 & 46.37 \\
J1243.1+3627 & - & BL Lac & HSP &  - &  - &  - &  - &  - &  - & -0.76 &  - &  - & 45.23 \\
J1248.2+5820 & - & BL Lac & ISP &  - &  - &  - &  - &  - &  - & -0.11 &  - &  - & 44.73 \\
J1253.2+5300 & - & BL Lac & LSP &  - &  - &  - &  - &  - &  - & -0.24 &  - &  - & 44.64 \\
J1256.1-0547 & 0.54 &   FSRQ & LSP & 12.87 &  0.48 &  1.38 & 11.49 & 36.92 & 42.97 &  - &  - &  - &  - \\
J1258.1+3233 & 0.81 &   FSRQ & LSP & 13.42 &  0.20 &  - &  - &  - &  - & -0.16 & 13.57 & 47.10 &  - \\
J1303.0+2435 & 0.99 & BL Lac & LSP & 14.03 &  0.30 &  - &  - &  - &  - &  0.48 & 13.55 & 42.55 &  - \\
J1308.7+3545 & 1.05 &   FSRQ & LSP & 13.37 &  0.25 &  - &  - &  - &  - &  0.81 & 12.56 & 40.20 &  - \\
J1310.6+3222 & 1.00 &   FSRQ & LSP & 13.53 & -0.12 &  1.19 & 12.34 & 38.17 & 42.99 &  1.12 & 12.41 & 38.65 & 43.13 \\
J1312.7+4828 & 0.64 &    AGN & LSP & 13.17 &  0.74 &  - &  - &  - &  - & -1.49 & 14.67 & 56.10 &  - \\
J1317.8+3429 & 1.05 &   FSRQ & LSP & 13.53 &  0.36 &  - &  - &  - &  - & -0.05 & 13.58 & 46.70 & 45.65 \\
J1326.8+2211 & 1.40 &   FSRQ & LSP & 13.43 &  0.47 &  1.33 & 12.10 & 37.56 & 42.46 &  1.39 & 12.04 & 37.13 & 42.34 \\
J1331.8+4718 & 0.67 &   FSRQ & LSP & 14.35 &  0.17 &  - &  - &  - &  - &  0.41 & 13.94 & 42.97 &  - \\
J1333.7+5057 & 1.36 &   FSRQ & ISP & 14.13 &  1.07 &  - &  - &  - &  - & -0.53 & 14.67 & 50.35 &  - \\
J1337.6-1257 & 0.54 &   FSRQ & LSP & 13.37 & -0.13 &  0.92 & 12.45 & 39.54 & 43.31 &  - &  - &  - &  - \\
J1345.6+4453 & 2.53 &   FSRQ & LSP & 13.55 &  1.04 &  - &  - &  - &  - &  0.63 & 12.92 & 43.15 &  - \\
J1350.8+3035 & 0.71 &   FSRQ & LSP & 13.87 &  0.15 &  - &  - &  - &  - & -0.18 & 14.06 & 47.08 &  - \\
J1357.6+7643 & 1.59 &   FSRQ & LSP & 13.08 &  0.44 &  - &  - &  - &  - &  0.23 & 12.85 & 45.02 &  - \\
J1359.0+5544 & 1.01 &   FSRQ & ISP & 13.53 &  1.35 &  - &  - &  - &  - &  0.10 & 13.43 & 45.89 &  - \\
J1416.0+1325 & 0.25 &  BCU I & LSP & 13.55 & -0.55 &  1.09 & 12.46 & 37.22 &  - &  - &  - &  - &  - \\
J1419.9+5425 & 0.15 & BL Lac & LSP & 13.98 & -0.54 &  0.71 & 13.27 & 39.51 & 42.37 &  - &  - &  - &  - \\
J1427.0+2347 & - & BL Lac & HSP &  - &  - &  - &  - &  - &  - & -0.62 &  - &  - & 45.45 \\
J1434.1+4203 & 1.24 &   FSRQ & LSP &  - &  - &  - &  - &  - &  - & -0.52 &  - &  - & 46.39 \\
J1436.8+2322 & 1.54 &   FSRQ & LSP & 13.49 & -0.17 &  - &  - &  - &  - &  0.75 & 12.74 & 41.35 &  - \\
J1438.7+3710 & 2.40 &   FSRQ & LSP & 12.95 &  0.86 &  - &  - &  - &  - &  0.66 & 12.29 & 42.43 &  - \\
J1443.9+2502 & 0.94 &   FSRQ &   - & 13.42 & -0.19 &  - &  - &  - &  - &  0.36 & 13.06 & 43.20 &  - \\
J1454.5+5124 & - & BL Lac & ISP &  - &  - &  - &  - &  - &  - & -0.93 &  - &  - & 47.13 \\
J1504.4+1029 & 1.84 &   FSRQ & LSP & 13.49 &  0.75 &  1.08 & 12.41 & 40.09 & 43.49 & -0.15 & 13.64 & 48.70 & 45.95 \\
J1506.1+3728 & 0.67 &   FSRQ & LSP & 12.89 &  0.35 &  - &  - &  - &  - &  0.23 & 12.66 & 44.19 &  - \\
J1512.8-0906 & 0.36 &   FSRQ & LSP & 13.58 &  0.96 &  1.22 & 12.36 & 37.89 & 42.48 &  - &  - &  - &  - \\
J1516.9+1926 & - & BL Lac & LSP &  - &  - &  - &  - &  - &  - &  0.66 &  - &  - & 42.75 \\
J1522.1+3144 & 1.49 &   FSRQ & LSP & 13.42 &  1.58 &  - &  - &  - &  - &  0.35 & 13.07 & 45.09 &  - \\
J1539.5+2746 & 2.19 &   FSRQ & ISP & 14.21 & -0.31 &  - &  - &  - &  - &  0.29 & 13.93 & 44.65 &  - \\
J1540.8+1449 & 0.61 & BL Lac & LSP & 13.58 & -0.81 &  0.63 & 12.94 & 41.32 & 43.99 &  - &  - &  - &  - \\
  \noalign{\smallskip}\hline
\end{tabular}
\ec
\end{table}

\setcounter{table}{0}
\begin{table}
\bc
\caption{continued.}
\setlength{\tabcolsep}{3pt}
\small
 \begin{tabular}{cccccccccccccc}
  \hline\noalign{\smallskip}
3FGL name  & z    & opt.     &  SED    &
 $\nu_{S,p}$    &  $CD$    &  $\delta_{var}$   &
 $\nu_{S,int}$     &  $L_{bol}$      &  $P_{j,int}$   &
 $\delta_{eq}$   &  $\nu_{S,int}$     &  $L_{bol}$    &  $P_{j,int}$ \\
  \hline\noalign{\smallskip}
J1542.9+6129 & - & BL Lac & ISP &  - &  - &  - &  - &  - &  - & -0.32 &  - &  - & 43.83 \\
J1553.5+1256 & 1.31 &   FSRQ & ISP & 13.80 &  0.15 &  - &  - &  - &  - & -0.62 & 14.42 & 51.10 & 46.75 \\
J1604.6+5714 & 0.72 &   FSRQ & ISP & 13.70 &  0.41 &  - &  - &  - &  - & -0.69 & 14.39 & 50.93 &  - \\
J1607.0+1551 & 0.50 &   FSRQ & LSP & 13.36 &  0.17 &  - &  - &  - &  - & -0.20 & 13.56 & 46.84 & 45.21 \\
J1608.6+1029 & 1.23 &   FSRQ & LSP & 13.39 &  0.98 &  1.40 & 11.99 & 37.53 & 42.49 & -0.08 & 13.47 & 47.90 & 45.46 \\
J1613.8+3410 & 1.40 &   FSRQ & LSP & 13.43 & -0.80 &  1.14 & 12.29 & 38.95 & 43.03 & -0.15 & 13.57 & 47.93 & 45.59 \\
J1630.6+8232 & 0.02 &    RDG & LSP & 14.29 & -0.94 &  - &  - &  - &  - & -0.55 & 14.84 & 46.59 &  - \\
J1635.2+3809 & 1.81 &   FSRQ & LSP & 13.45 &  0.74 &  1.33 & 12.12 & 38.44 & 42.96 &  - &  - &  - &  - \\
J1637.7+4715 & 0.74 &   FSRQ & LSP & 13.06 &  0.32 &  - &  - &  - &  - &  0.18 & 12.88 & 44.75 & 45.06 \\
J1637.9+5719 & 0.75 &   FSRQ & ISP & 13.47 &  1.37 &  1.15 & 12.33 & 39.16 &  - &  - &  - &  - &  - \\
J1640.9+1142 & 0.08 &  BCU I & ISP & 13.93 & -0.60 &  - &  - &  - &  - & -0.82 & 14.75 & 49.17 &  - \\
J1642.9+3950 & 0.59 &   FSRQ & LSP & 13.60 & -0.29 &  0.89 & 12.71 & 40.26 & 43.96 &  - &  - &  - &  - \\
J1647.4+4950 & 0.05 &  BCU I & LSP & 14.68 &  0.22 &  - &  - &  - &  - & -0.39 & 15.07 & 46.25 &  - \\
J1656.9+6008 & 0.62 &   FSRQ & ISP & 13.63 &  0.82 &  - &  - &  - &  - &  0.16 & 13.47 & 44.93 &  - \\
J1700.1+6829 & 0.30 &   FSRQ & LSP & 13.36 &  0.87 &  - &  - &  - &  - &  0.16 & 13.20 & 44.15 & 44.04 \\
J1709.6+4318 & 1.03 &   FSRQ & LSP & 13.97 &  0.79 &  - &  - &  - &  - & -0.36 & 14.33 & 48.88 &  - \\
J1719.2+1744 & 0.14 & BL Lac & LSP & 13.32 & -0.27 &  - &  - &  - &  - &  0.52 & 12.79 & 40.32 & 42.51 \\
J1722.7+1014 & 0.73 &   FSRQ & LSP & 13.53 &  0.17 &  - &  - &  - &  - & -0.41 & 13.94 & 48.78 &  - \\
J1727.1+4531 & 0.72 &   FSRQ & LSP & 13.70 &  0.52 &  - &  - &  - &  - &  1.53 & 12.18 & 35.71 & 42.04 \\
J1728.3+5013 & 0.06 & BL Lac & HSP & 15.98 & -1.42 &  - &  - &  - &  - & -0.86 & 16.84 & 49.94 & 45.00 \\
J1728.5+0428 & 0.29 &   FSRQ & LSP & 13.82 &  0.62 &  0.58 & 13.24 & 41.49 & 43.02 &  - &  - &  - &  - \\
J1730.6+3711 & 0.20 & BL Lac & ISP & 14.68 & -0.80 &  - &  - &  - &  - & -1.42 & 16.10 & 54.18 &  - \\
J1733.0-1305 & 0.90 &   FSRQ & LSP & 12.44 &  0.36 &  1.03 & 11.41 & 39.40 & 43.66 &  - &  - &  - &  - \\
J1734.3+3858 & 0.98 &   FSRQ & LSP & 13.27 &  0.31 &  - &  - &  - &  - &  0.39 & 12.88 & 43.56 &  - \\
J1740.3+5211 & 1.38 &   FSRQ & LSP & 13.91 &  0.32 &  1.42 & 12.48 & 36.98 & 42.52 &  - &  - &  - &  - \\
J1742.2+5947 & - & BL Lac & ISP &  - &  - &  - &  - &  - &  - & -0.50 &  - &  - & 44.38 \\
J1743.9+1934 & 0.08 & BL Lac & HSP & 15.23 & -1.28 &  - &  - &  - &  - & -0.65 & 15.87 & 48.64 & 45.11 \\
J1744.3-0353 & 1.06 &   FSRQ & LSP & 13.26 & -0.57 &  1.29 & 11.97 & 37.13 &  - &  - &  - &  - &  - \\
J1748.6+7005 & 0.77 & BL Lac & LSP & 14.57 & -0.49 &  - &  - &  - &  - & -0.24 & 14.81 & 47.98 & 45.33 \\
J1749.1+4322 & - & BL Lac & LSP &  - &  - &  - &  - &  - &  - &  0.11 &  - &  - & 43.81 \\
J1751.5+0939 & 0.32 & BL Lac & LSP & 13.85 &  0.03 &  1.08 & 12.77 & 37.67 & 42.70 &  - &  - &  - &  - \\
J1800.5+7827 & 0.68 & BL Lac & LSP & 13.76 & -0.23 &  1.09 & 12.67 & 38.67 & 42.71 &  0.41 & 13.35 & 43.42 & 44.06 \\
J1806.7+6949 & 0.05 & BL Lac & ISP & 14.20 & -0.66 &  0.04 & 14.16 & 43.52 & 43.87 & -0.05 & 14.25 & 44.16 & 44.05 \\
J1813.6+3143 & 0.12 & BL Lac & ISP & 14.81 & -0.75 &  - &  - &  - &  - & -0.58 & 15.39 & 48.12 & 44.78 \\
J1824.2+5649 & 0.66 & BL Lac & LSP & 13.73 &  0.23 &  0.81 & 12.92 & 40.66 & 43.79 &  0.56 & 13.17 & 42.37 & 44.28 \\
J1829.6+4844 & 0.69 &   SSRQ & LSP & 13.32 & -0.89 &  0.76 & 12.56 & 41.36 &  - &  - &  - &  - &  - \\
J1848.4+3216 & 0.80 &   FSRQ & LSP & 13.28 &  0.77 &  - &  - &  - &  - & -0.15 & 13.43 & 47.31 &  - \\
J1849.2+6705 & 0.66 &   FSRQ & LSP & 13.84 &  0.25 &  - &  - &  - &  - &  0.89 & 12.95 & 39.76 & 43.62 \\
J1852.4+4856 & 1.25 &   FSRQ & LSP & 13.53 &  0.58 &  - &  - &  - &  - &  1.17 & 12.36 & 38.17 &  - \\
J2000.0+6509 & 0.05 & BL Lac & HSP & 16.19 & -0.84 &  - &  - &  - &  - & -0.34 & 16.53 & 46.35 &  - \\
J2001.8+7041 & 0.25 & BL Lac & HSP & 13.52 & -0.18 &  - &  - &  - &  - & -0.95 & 14.47 & 50.83 &  - \\
J2005.2+7752 & 0.34 & BL Lac & LSP & 14.07 & -0.60 &  1.12 & 12.94 & 37.51 & 42.08 &  0.29 & 13.78 & 43.36 & 43.75 \\
J2022.5+7612 & 0.59 & BL Lac & ISP & 14.51 & -0.28 &  - &  - &  - &  - &  0.38 & 14.14 & 43.24 &  - \\
J2031.8+1223 & 1.22 & BL Lac & LSP & 14.04 & -0.20 &  - &  - &  - &  - &  0.98 & 13.06 & 39.85 &  - \\
J2035.3+1055 & 0.60 &   FSRQ & ISP & 13.90 &  0.22 &  - &  - &  - &  - &  0.43 & 13.47 & 43.13 & 44.01 \\
  \noalign{\smallskip}\hline
\end{tabular}
\ec
\end{table}

\setcounter{table}{0}
\begin{table}
\bc
\caption{continued.}
\setlength{\tabcolsep}{3pt}
\small
 \begin{tabular}{cccccccccccccc}
  \hline\noalign{\smallskip}
3FGL name  & z    & opt.     &  SED    &
 $\nu_{S,p}$    &  $CD$    &  $\delta_{var}$   &
 $\nu_{S,int}$     &  $L_{bol}$      &  $P_{j,int}$   &
 $\delta_{eq}$   &  $\nu_{S,int}$     &  $L_{bol}$    &  $P_{j,int}$ \\
  \hline\noalign{\smallskip}
J2115.4+2933 & 1.51 &   FSRQ & LSP & 13.33 &  0.25 &  - &  - &  - &  - & -0.84 & 14.17 & 52.54 &  - \\
J2116.1+3339 & 1.60 & BL Lac & HSP & 15.35 & -0.17 &  - &  - &  - &  - & -1.05 & 16.40 & 54.14 &  - \\
J2121.0+1901 & 2.18 &   FSRQ & ISP & 13.33 &  0.11 &  - &  - &  - &  - & -0.35 & 13.68 & 49.32 &  - \\
J2123.6+0533 & 1.94 &   FSRQ & LSP & 13.98 & -0.92 &  1.18 & 12.79 & 38.55 &  - &  - &  - &  - &  - \\
J2143.5+1744 & 0.21 &   FSRQ & ISP & 14.23 &  0.78 &  - &  - &  - &  - &  0.48 & 13.75 & 42.18 & 43.41 \\
J2152.4+1735 & 0.87 & BL Lac & LSP & 13.83 & -1.27 &  - &  - &  - &  - & -0.52 & 14.35 & 49.57 &  - \\
J2202.7+4217 & 0.07 & BL Lac & LSP & 14.10 & -0.36 &  0.86 & 13.24 & 38.59 & 41.72 &  - &  - &  - &  - \\
J2203.4+1725 & 1.08 &   FSRQ & LSP & 14.05 &  0.18 &  - &  - &  - &  - &  0.12 & 13.93 & 45.75 & 45.26 \\
J2203.7+3143 & 0.29 &   FSRQ & LSP & 14.52 &  2.25 &  0.83 & 13.70 & 42.23 &  - &  - &  - &  - &  - \\
J2212.0+2355 & 1.13 &   FSRQ & LSP & 13.31 & -0.10 &  - &  - &  - &  - &  0.41 & 12.89 & 43.16 &  - \\
J2217.0+2421 & 0.50 & BL Lac & LSP & 14.28 & -0.68 &  - &  - &  - &  - &  0.82 & 13.46 & 39.83 & 43.14 \\
J2225.8-0454 & 1.40 &   FSRQ & LSP & 13.24 & -0.44 &  1.20 & 12.04 & 38.70 & 43.88 &  - &  - &  - &  - \\
J2229.7-0833 & 1.56 &   FSRQ & LSP & 13.89 &  1.00 &  1.20 & 12.69 & 39.26 & 42.77 &  - &  - &  - &  - \\
J2232.5+1143 & 1.04 &   FSRQ & LSP & 13.42 &  0.41 &  1.19 & 12.23 & 38.99 & 43.33 &  - &  - &  - &  - \\
J2236.3+2829 & 0.79 & BL Lac & LSP & 14.08 & -0.13 &  0.78 & 13.30 & 40.91 &  - &  0.72 & 13.36 & 41.32 &  - \\
J2250.1+3825 & 0.12 & BL Lac &   - & 15.54 & -0.90 &  - &  - &  - &  - & -0.97 & 16.51 & 50.96 &  - \\
J2251.9+4031 & 0.23 & BL Lac & ISP & 14.25 &  0.03 &  - &  - &  - &  - & -0.74 & 14.99 & 49.54 &  - \\
J2254.0+1608 & 0.86 &   FSRQ & LSP & 13.64 &  0.76 &  1.52 & 12.12 & 37.12 & 42.82 &  1.07 & 12.57 & 40.26 & 43.71 \\
J2311.0+3425 & 1.82 &   FSRQ & LSP & 13.72 &  0.47 &  - &  - &  - &  - &  0.06 & 13.66 & 46.84 & 45.66 \\
J2321.9+2732 & 1.25 &   FSRQ & LSP & 13.63 & -0.58 &  - &  - &  - &  - & -0.07 & 13.70 & 47.00 &  - \\
J2321.9+3204 & 1.49 &   FSRQ & LSP & 13.59 &  0.36 &  - &  - &  - &  - &  0.67 & 12.92 & 41.76 &  - \\
J2322.5+3436 & 0.10 & BL Lac & HSP & 15.27 & -0.27 &  - &  - &  - &  - & -0.87 & 16.14 & 50.06 &  - \\
  \noalign{\smallskip}\hline
\end{tabular}
\ec
%% place \tablecomments and \tablerefs below \end{center| and \end{center}:
%% you may leave the table-width parameter to editors or set to your actual size
\tablecomments{0.86\textwidth}{Column 1 is the source name in the third Fermi-LAT catalog (3FGL,~\citealt{2015ApJS..218...23A}). Columns 2-4 give the redshift, optical type and the SED classification in the 3LAC. Column 5 and Column 6 are the k-corrected synchrotron peak frequency in unit of Hz, and Compton Dominance. Column 7 is the Doppler factor estimated from radio variability. Columns 8-10 are the Doppler-corrected intrinsic synchrotron peak frequency in unit of Hz, intrinsic luminosity in unit of erg s$^{-1}$, and intrinsic jet power in unit of erg s$^{-1}$ calculated with $\delta_{var}$. Column 11 is the Doppler factor derived from VLBI observations. Columns 12-14 are the Doppler-corrected intrinsic synchrotron peak frequency in unit of Hz, intrinsic luminosity in unit of erg s$^{-1}$, and intrinsic jet power in unit of erg s$^{-1}$ calculated with $\delta_{eq}$. All the values except redshift are in logarithmic space. }
\end{table}

\label{lastpage}


\begin{thebibliography}{37}
\providecommand\natexlab[1]{#1}
\providecommand\JournalTitle[1]{#1}

\bibitem[{Abdo} {et~al.}(2010)]{2010ApJ...716...30A}
{Abdo}, A.~A., {Ackermann}, M., {Agudo}, I., {et~al.} 2010, ApJ, 716, 30

\bibitem[{Acero} {et~al.}(2015)]{2015ApJS..218...23A}
{Acero}, F., {Ackermann}, M., {Ajello}, M., {et~al.} 2015, \apjs, 218, 23

\bibitem[{Ackermann} {et~al.}(2015)]{2015ApJ...810...14A}
{Ackermann}, M., {Ajello}, M., {Atwood}, W.~B., {et~al.} 2015, \apj, 810, 14

\bibitem[{Akritas} \& {Siebert}(1996)]{1996MNRAS.278..919A}
{Akritas}, M.~G., \& {Siebert}, J. 1996, \mnras, 278, 919

\bibitem[{Celotti} \& {Ghisellini}(2008)]{2008MNRAS.385..283C}
{Celotti}, A., \& {Ghisellini}, G. 2008, \mnras, 385, 283

\bibitem[{Chai} {et~al.}(2012)]{2012ApJ...759..114C}
{Chai}, B., {Cao}, X., \& {Gu}, M. 2012, \apj, 759, 114

\bibitem[{Chen}(2014)]{2014ApJ...788..179C}
{Chen}, L. 2014, ApJ, 788, 179

\bibitem[{Chen} \& {Bai}(2011)]{2011ApJ...735..108C}
{Chen}, L., \& {Bai}, J.~M. 2011, ApJ, 735, 108

\bibitem[{Finke}(2013)]{2013ApJ...763..134F}
{Finke}, J.~D. 2013, ApJ, 763, 134

\bibitem[{Fossati} {et~al.}(1998)]{1998MNRAS.299..433F}
{Fossati}, G., {Maraschi}, L., {Celotti}, A., {Comastri}, A., \& {Ghisellini},
  G. 1998, MNRAS, 299, 433

\bibitem[{Ghisellini} {et~al.}(1998)]{1998MNRAS.301..451G}
{Ghisellini}, G., {Celotti}, A., {Fossati}, G., {Maraschi}, L., \& {Comastri},
  A. 1998, MNRAS, 301, 451

\bibitem[{Ghisellini} {et~al.}(2009)]{2009MNRAS.396L.105G}
{Ghisellini}, G., {Maraschi}, L., \& {Tavecchio}, F. 2009, MNRAS, 396, L105

\bibitem[{Ghisellini} {et~al.}(1993)]{1993ApJ...407...65G}
{Ghisellini}, G., {Padovani}, P., {Celotti}, A., \& {Maraschi}, L. 1993, \apj,
  407, 65

\bibitem[{Ghisellini} \& {Tavecchio}(2008)]{2008MNRAS.387.1669G}
{Ghisellini}, G., \& {Tavecchio}, F. 2008, MNRAS, 387, 1669

\bibitem[{Ghisellini} {et~al.}(2011)]{2011MNRAS.414.2674G}
{Ghisellini}, G., {Tavecchio}, F., {Foschini}, L., \& {Ghirlanda}, G. 2011,
  MNRAS, 414, 2674

\bibitem[{Giommi} {et~al.}(2012)]{2012MNRAS.420.2899G}
{Giommi}, P., {Padovani}, P., {Polenta}, G., {et~al.} 2012, MNRAS, 420, 2899

\bibitem[{Hovatta} {et~al.}(2009)]{2009A&A...494..527H}
{Hovatta}, T., {Valtaoja}, E., {Tornikoski}, M., \& {L{\"a}hteenm{\"a}ki}, A.
  2009, \aap, 494, 527

\bibitem[{Komatsu} {et~al.}(2009)]{2009ApJS..180..330K}
{Komatsu}, E., {Dunkley}, J., {Nolta}, M.~R., {et~al.} 2009, ApJS, 180, 330

\bibitem[{Kovalev} {et~al.}(2009)]{2009ApJ...696L..17K}
{Kovalev}, Y.~Y., {Aller}, H.~D., {Aller}, M.~F., {et~al.} 2009, \apjl, 696,
  L17

\bibitem[{L{\"a}hteenm{\"a}ki} \& {Valtaoja}(1999)]{1999ApJ...521..493L}
{L{\"a}hteenm{\"a}ki}, A., \& {Valtaoja}, E. 1999, \apj, 521, 493

\bibitem[{Linford} {et~al.}(2012)]{2012ApJ...744..177L}
{Linford}, J.~D., {Taylor}, G.~B., {Romani}, R.~W., {et~al.} 2012, ApJ, 744,
  177

\bibitem[{Lister} {et~al.}(2015)]{2015ApJ...810L...9L}
{Lister}, M.~L., {Aller}, M.~F., {Aller}, H.~D., {et~al.} 2015, \apjl, 810, L9

\bibitem[{Lister} {et~al.}(2009{\natexlab{a}})]{2009ApJ...696L..22L}
{Lister}, M.~L., {Homan}, D.~C., {Kadler}, M., {et~al.} 2009{\natexlab{a}},
  \apjl, 696, L22

\bibitem[{Lister} {et~al.}(2009{\natexlab{b}})]{2009AJ....138.1874L}
{Lister}, M.~L., {Cohen}, M.~H., {Homan}, D.~C., {et~al.} 2009{\natexlab{b}},
  AJ, 138, 1874

\bibitem[{Meyer} {et~al.}(2011)]{2011ApJ...740...98M}
{Meyer}, E.~T., {Fossati}, G., {Georganopoulos}, M., \& {Lister}, M.~L. 2011,
  ApJ, 740, 98

\bibitem[{Meyer} {et~al.}(2012)]{2012ApJ...752L...4M}
{Meyer}, E.~T., {Fossati}, G., {Georganopoulos}, M., \& {Lister}, M.~L. 2012,
  ApJL, 752, L4

\bibitem[{Nemmen} {et~al.}(2012)]{2012Sci...338.1445N}
{Nemmen}, R.~S., {Georganopoulos}, M., {Guiriec}, S., {et~al.} 2012, Science,
  338, 1445

\bibitem[{Nieppola} {et~al.}(2008)]{2008A&A...488..867N}
{Nieppola}, E., {Valtaoja}, E., {Tornikoski}, M., {Hovatta}, T., \&
  {Kotiranta}, M. 2008, A\&A, 488, 867

\bibitem[{Padovani} {et~al.}(2012)]{2012MNRAS.422L..48P}
{Padovani}, P., {Giommi}, P., \& {Rau}, A. 2012, MNRAS, 422, L48

\bibitem[{Potter} \& {Cotter}(2013{\natexlab{a}})]{2013MNRAS.431.1840P}
{Potter}, W.~J., \& {Cotter}, G. 2013{\natexlab{a}}, \mnras, 431, 1840

\bibitem[{Potter} \& {Cotter}(2013{\natexlab{b}})]{2013MNRAS.436..304P}
{Potter}, W.~J., \& {Cotter}, G. 2013{\natexlab{b}}, \mnras, 436, 304

\bibitem[{Readhead}(1994)]{1994ApJ...426...51R}
{Readhead}, A.~C.~S. 1994, ApJ, 426, 51

\bibitem[{Richards} {et~al.}(2011)]{2011ApJS..194...29R}
{Richards}, J.~L., {Max-Moerbeck}, W., {Pavlidou}, V., {et~al.} 2011, \apjs,
  194, 29

\bibitem[{Savolainen} {et~al.}(2010)]{2010A&A...512A..24S}
{Savolainen}, T., {Homan}, D.~C., {Hovatta}, T., {et~al.} 2010, A\&A, 512, A24

\bibitem[{Sikora} {et~al.}(2009)]{2009ApJ...704...38S}
{Sikora}, M., {Stawarz}, {\L}., {Moderski}, R., {Nalewajko}, K., \& {Madejski},
  G.~M. 2009, \apj, 704, 38

\bibitem[{Tavecchio} {et~al.}(1998)]{1998ApJ...509..608T}
{Tavecchio}, F., {Maraschi}, L., \& {Ghisellini}, G. 1998, \apj, 509, 608

\bibitem[{Urry} \& {Padovani}(1995)]{1995PASP..107..803U}
{Urry}, C.~M., \& {Padovani}, P. 1995, \pasp, 107, 803

\end{thebibliography}
\end{document}